\documentclass[12pt]{iopart}
\usepackage{graphicx}% Include figure files
\usepackage{dcolumn}% Align table columns on decimal point

\usepackage{epstopdf}
\begin{document}

\title{Review of annealing effects and superconductivity in Fe$_{1+y}$Te$_{1-x}$Se$_x$ superconductors}

\author{Yue Sun,$^{1,2,3*}$ Zhixiang Shi$^{1\star}$ and Tsuyoshi Tamegai$^3$}

\address{
$^1$School of Physics and Key Laboratory of MEMS of the Ministry of Education, Southeast University, Nanjing 211189, China\\
$^2$Department of Physics and Mathematics, Aoyama Gakuin University, Sagamihara 252-5258, Japan\\
$^3$Department of Applied Physics, The University of Tokyo, Bunkyo-ku, Tokyo 113-8656, Japan}

\ead{$^ \star$zxshi@seu.edu.cn; $^*$sunyue@phys.aoyama.ac.jp}
\vspace{10pt}

%\begin{indented}
%\item[]March 2019
%\end{indented}

\begin{abstract}
Among iron-based superconductors, Fe$_{1+y}$Te$_{1-x}$Se$_x$ is unique in their structural simplicity, consisting of only FeTe/Se layers, which is favorable for probing the mechanism of superconductivity. Recently, a topological surface superconductivity as well as the Majorana Fermions has been observed, which makes Fe$_{1+y}$Te$_{1-x}$Se$_x$ the first high temperature topological superconductor. Thus, Fe$_{1+y}$Te$_{1-x}$Se$_x$ is unique for study both the high temperature and topological superconductivities. Since large size single crystal of Fe$_{1+y}$Te$_{1-x}$Se$_x$ can be easily grown, many researches have been performed. However, a large part of the reported results are under controversy, or different from groups, including those related to the basic properties like resistivity, susceptibility, Hall effect, gap structure, phase diagram, etc. These controversies are believed to come from the sample-dependent Fe nonstoichiometries, which originate from the partial occupation of the second Fe site (excess Fe site, expressed as $y$ in the formula of Fe$_{1+y}$Te$_{1-x}$Se$_x$) in the Te/Se layer. The excess Fe with valence near Fe$^+$ will provide electron doping into the system. Meanwhile, the excess Fe is also strongly magnetic, which provides local moments that interact with adjacent Fe layers. The magnetic moment from excess Fe will act as a paring breaker and also localize the charge carriers. Thus, the existence of excess Fe complicates the study of Fe$_{1+y}$Te$_{1-x}$Se$_x$ from superconducting to normal state properties. Removing the excess Fe is essential to probe the intrinsic properties and mechanism of superconductivity, as well as to the applications of Fe$_{1+y}$Te$_{1-x}$Se$_x$ compounds. In this topical review, we present an overview of the reported annealing methods, and propose the effective approaches to remove excess Fe in Fe$_{1+y}$Te$_{1-x}$Se$_x$. Furthermore, we discuss the mechanism of annealing based on the evolutions of structure, composition, and morphology with annealing. Moreover, we also review the annealing effects on the normal state and superconducting properties, including the magnetism, transport properties, band structure, $T_{\rm{c}}$, phase diagram, upper critical field, anisotropy, critical current density, gap structure, and superconducting pairing. This review presents not only the optimal way to prepare crystals without excess Fe, but also the intrinsic properties of Fe$_{1+y}$Te$_{1-x}$Se$_x$ without the influence of excess Fe.

\end{abstract}
%\tableofcontents
%\newpage
%\maketitle

\section{Introduction}
As a member of the iron-based superconductors (IBSs), iron chalcogenide FeSe was first reported to be superconductors since 2008 by Hsu $et$ $al$. \cite{HsuFongChiFeSediscovery}. Although the initial superconducting transition temperature, $T_{\rm{c}}$, of FeSe is only $\sim$ 9 K, it was increased up to 14 K with an appropriate Te substitution \cite{SalesPRB} and 37 K under high pressure \cite{MedvedevNatMat}. Furthermore, by intercalating spacer layers between adjacent FeSe layers, $T_{\rm{c}}$ has been enhanced over 40 K, such as the Li$_x$(NH$_2$)$_y$(NH$_3$)$_{1-y}$Fe$_2$Se$_2$ \cite{BurrardNatMat}, Li$_{1-x}$Fe$_x$(OH)Fe$_{1-y}$Se \cite{LiFeOHSeNatMater}, and Na(C$_2$H$_8$N$_2$)$_{1.5}$Fe$_2$Se$_2$ \cite{HayashiInorg}. Moreover, gate-induced carrier doping using the electric double layer (EDL) was reported to increase the $T_{\rm{c}}$ of FeSe flakes to 48 K. More interestingly, angle-resolved photoemission spectroscopy (ARPES) revealed unexpectedly large superconducting gap $\sim$ 19 meV in a monolayer FeSe, suggesting a $T_{\rm{c}}$ as high as 65 K \cite{HeNatMat}, and a sign of $T_{\rm{c}}$ even above 100 K was also reported \cite{GeNatMat}. Thus, $T_{\rm{c}}$ of the iron chalcogenides can be easily tuned by doping, pressure, intercalating, or even reducing the dimensionality, which is fascinating for the study of superconductivity (SC).    

Among iron chalcogenides, Fe$_{1+y}$Te$_{1-x}$Se$_x$ is unique in structural simplicity, composing of only iron-chalcogenide layers, which is favorable for probing the mechanism of superconductivity. Band structure calculation \cite{SubediPRB} and ARPES \cite{XiaPRL} results show that the Fermi surface of FeTe$_{1-x}$Se$_x$ is characterized by hole bands around $\Gamma$ point and electron bands around $M$ point, which is similar to iron pnictides. Moreover, the competition between magnetism and superconductivity, similar to iron pnictides, is also observed in iron chalcogenides \cite{LiuNatMat}. The parent compound Fe$_{1+y}$Te is not superconducting, but exhibits a spin-density wave (SDW) ground state. With Se doping, superconductivity emerges and $T_{\rm{c}}$ goes up to 14 K accompanied by the suppression of SDW. One the other hand, iron chalcogenides also manifest some differences compared with iron pnictides. In particular, Fe$_{1+y}$Te$_{1-x}$Se$_x$ exhibits a bi-collinear antiferromagnetic ordering and it can be either commensurate or incommensurate depending on the Fe content \cite{BaoWeiPRL,ShiliangLiPhysRevB.79.054503}, different from the common collinear commensurate antiferromagnetic ordering observed in iron pnictides. 

In band structure, an extremely small Fermi energy is observed in Fe$_{1+y}$Te$_{1-x}$Se$_x$ by ARPES measurements, which indicates that it is in the crossover regime from Bardeen-Cooper-Schrieffer (BCS) to Bose-Einstein-condensation (BEC) \cite{Lubashevskynatphy}. Recently, a topological surface superconductivity \cite{ZhangARPESScience,ZhangPengNatPhy} as well as the Majorana Fermions has been observed \cite{WangMajoranaScience,MachdiaarXiv}, which makes Fe$_{1+y}$Te$_{1-x}$Se$_x$ the first high temperature topological superconductor. Those features make Fe$_{1+y}$Te$_{1-x}$Se$_x$ one of the most attractive compounds in the study of condensed matter physics. On the other hand, its high upper critical magnetic field ($\sim$ 50 Tesla) and the less toxic nature compared with iron pnictides suggest that Fe$_{1+y}$Te$_{1-x}$Se$_x$ are also favorable to applications. Until now, the tapes with transport $J_{\rm{c}}$ over 10$^6$ A/cm$^2$ under self-field and over 10$^5$ A/cm$^2$ under 30 T at 4.2 K have already been fabricated \cite{SiWeidongNatComm}.   

Although many efforts have been done on both fundamental and application studies of Fe$_{1+y}$Te$_{1-x}$Se$_x$, some fundamental properties are still under controversy. For the resistivity, both metallic and nonmetallic behaviors were reported, and the absolute value just above $T_{\rm{c}}$ has a spread from 200 to 1500 $\mu\Omega$cm \cite{SalesPRB,TaenPRB,NojiFeTeSeannealing,LiuPRB,LiuSUST,SunPRB,DongPRB}. For the Hall coefficient, conflicting low-temperature behaviors have been observed in crystals with nominally the same amount of Se \cite{LiuPRB,LiuSUST,SunPRB}. The reported critical current density, $J_{\rm{c}}$, measured at the same situation (such as 5 K, zero field), ranging from 10$^4$ A/cm$^2$ to over 10$^5$ A/cm$^2$ in the optimally-doped crystals \cite{TaenPRB,BonuraJcPhysRevB.85.134532,DasJcPhysRevB.84.214526,YadavJcNJP}. In the phase diagram, although Liu $et$ $al$. reported that bulk superconductivity resides only in the region of Se level higher than 29\% \cite{LiuNatMat}, it was observed in the Se doping level between 10 and 50\% by Noji $et$ $al$. \cite{NojiJPSJPhaseD}. For the gap structure, two ARPES measurements support isotropic superconducting gap \cite{ZhangARPESScience,MiaoPRB}, while anisotropic gap structure was suggested by another ARPES report \cite{OkazakiARPESPhysRevLett.109.237011} and angle-resolved specific heat measurement \cite{ZengNatCommun}. 

These controversies are believed to come from the sample-dependent Fe nonstoichiometries, which originate from the partial occupation of the second Fe site (excess Fe site, expressed as $y$ in the formula of Fe$_{1+y}$Te$_{1-x}$Se$_x$) in the Te/Se layer \cite{BaoWeiPRL,BendelePRB}. The excess Fe with valence near Fe$^+$ will provide electron doping into the 11 system. The excess Fe is also strongly magnetic, which provides local moments that interact with the adjacent Fe layers \cite{ZhangPRB}. The magnetic moment from excess Fe will act as a pair breaker and also localize the charge carriers \cite{LiuPRB,SunPRB}. In the study of the topological SC of Fe$_{1+y}$Te$_{1-x}$Se$_x$, the excess Fe obscures the surface state in the ARPES measurements \cite{Zhangpengprivatecoom}. The excess Fe will also lift up the gap-bottom of the tunneling spectrum in STM measurements, and contributes a robust zero-energy bound state, which disturbs the observation of possible Majorana Fermions \cite{YinSTMNatPhy}. Thus, the existence of excess Fe complicates the study of Fe$_{1+y}$Te$_{1-x}$Se$_x$ from normal state to superconducting properties, as well as fundamental and applied research. 

Unfortunately, the excess Fe is almost unavoidable in the current growth technique employing slow cooling. To probe the intrinsic properties of Fe$_{1+y}$Te$_{1-x}$Se$_x$, many groups have tried different methods to remove the excess Fe, including annealing in various atmospheres, deintercalating by immersing crystals in liquids, and the electro-chemical methods. Some methods are effective to totally remove the excess Fe, and induce bulk SC, while others only partially reduce the excess Fe or even invalid. Hence, the reported properties of the annealed (including the deintercalated and electro-chemical treated) crystals from different groups vary considerably. 

In this review, we first summarize all the reported annealing methods, and propose the most effective approaches to remove the excess Fe in Fe$_{1+y}$Te$_{1-x}$Se$_x$ (section 2.1-2.2). We discuss the mechanism of annealing based on the evolutions of structure, composition, and morphology with annealing (section 2.3-2.4). Then, we systematically review the annealing effects based on the evolution of physical properties before and after annealing. In section 3, we discuss the annealing effect on the normal state properties, including the magnetism, transport properties, and the band structure. In section 4, we discuss the annealing effects on the superconducting properties, including the $T_{\rm{c}}$, phase diagram, upper critical field, anisotropy, critical current density, the gap structure, and superconducting pairing. Finally, we conclude this review with summary and perspective in section 5.

\section{Experiments}

\subsection{Crystal growth}
FeTe$_{1-x}$Se$_x$ ($x$ = 0 $\sim$ 0.4) single crystal can be successfully grown using standard melting techniques, including the self-flux method with slow cooling \cite{SunSUST} (in some paper it was called as Bridgman method) and optical zone-melting \cite{Wenjinshengreview}. The self-flux method is much easier and often used to grow FeTe$_{1-x}$Se$_x$ single crystal. Single crystals grown by the melting methods usually unavoidably contains excess Fe residing in the interstitial sites of the Te/Se layer, which may be preferable in stabilizing the crystal structure during the growth of single crystals \cite{Wenjinshengreview}. In this paper, we use $y$ in the Fe$_{1+y}$Te$_{1-x}$Se$_x$ to represent the amount of excess Fe. In detail, high purity Fe (99.99\%), Te (99.999\%), Se (99.999\%) grains with stoichiometric compositions of FeTe$_{1-x}$Se$_x$ (more than 10 g) were loaded into a small quartz tube with $d_1$ $\sim$ 10 mm $\phi$, evacuated, and sealed. Then the small tube was sealed into a second evacuated quartz tube with $d_2$ $\sim$ 20 mm since the small tube often cracks during the cooling precess. The whole assembly was heated up to 1070$^\circ$C and kept for 36 h, followed by slow cooling down to 710$^\circ$C with a rate slower than $\sim$ 6$^\circ$C/h. After that, the assembly was cooled down to room temperature by shutting down the furnace. Large size single crystals reaching centimeter-scale with a mirror-like surface (corresponding to the $ab$-plane) as shown in Fig. 1(a) can be obtained. The excess Fe is indicated by the orange ball in the crystal structure shown in Fig. 1(b), whose amount depends on the growth condition, and was reported to be as large as $\sim$ 15\% \cite{BaoWeiPRL,SunSciRep}.          

\begin{figure}\center	
	\includegraphics[width=8.5cm]{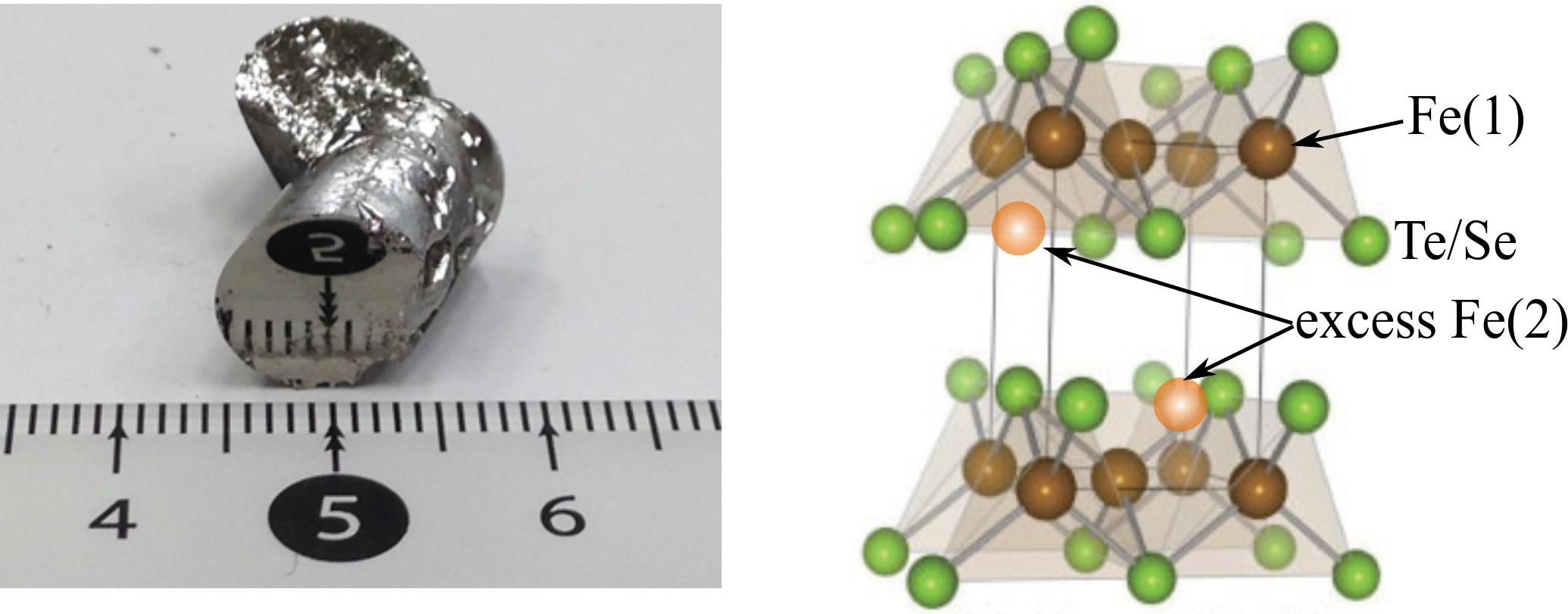}\\
	\caption{(a) Photograph of the as-grown FeTe$_{0.6}$Se$_{0.4}$ single crystal \cite{SunSUSTevolution}. (b) Crystal structure of FeTe$_{1-x}$Se$_x$ \cite{FujitsuHosonoreview}. The orange ball represents the excess Fe.}\label{}
\end{figure}

\subsection{Annealing in various atmospheres and liquids}

The first attempt to anneal Fe$_{1+y}$Te$_{1-x}$Se$_x$ was made by Taen $et$ $al$. \cite{TaenPRB}. They found that bulk superconductivity with $T_{\rm{c}}$ $\sim$ 14 K can be induced by vacuum annealing the non-superconducting Fe$_{1+y}$Te$_{0.61}$Se$_{0.39}$ at 400$^\circ$C for more than 10 days. After that, vacuum annealing was also applied to remove the excess Fe in Fe$_{1+y}$Te$_{1-x}$Se$_x$ single crystals with $x$ ranging from 0 to 0.5 \cite{NojiFeTeSeannealing,KomiyavacumnannealJPSJ.82.064710}. Then, N$_2$ annealing was also reported to be effective to remove excess Fe \cite{HuN2HNO3annealSUST}. However, it was later confirmed that the vacuum and N$_2$ annealing have no effect to the excess Fe, and the reported \textquotedblleft vacuum\textquotedblright annealing effect has been proved to be from the effect of residual small amount of O$_2$ \cite{SunSUST}, which will be discussed later. Due to the effect of O$_2$, crystals annealed in the air condition \cite{DongPRB} also show the improvement of superconductivity. Therefore, we have to be very careful about the vacuum level during the study of the annealing effect as well as searching for the optimal annealing condition.  

\begin{figure}\center	
	\includegraphics[width=8.5cm]{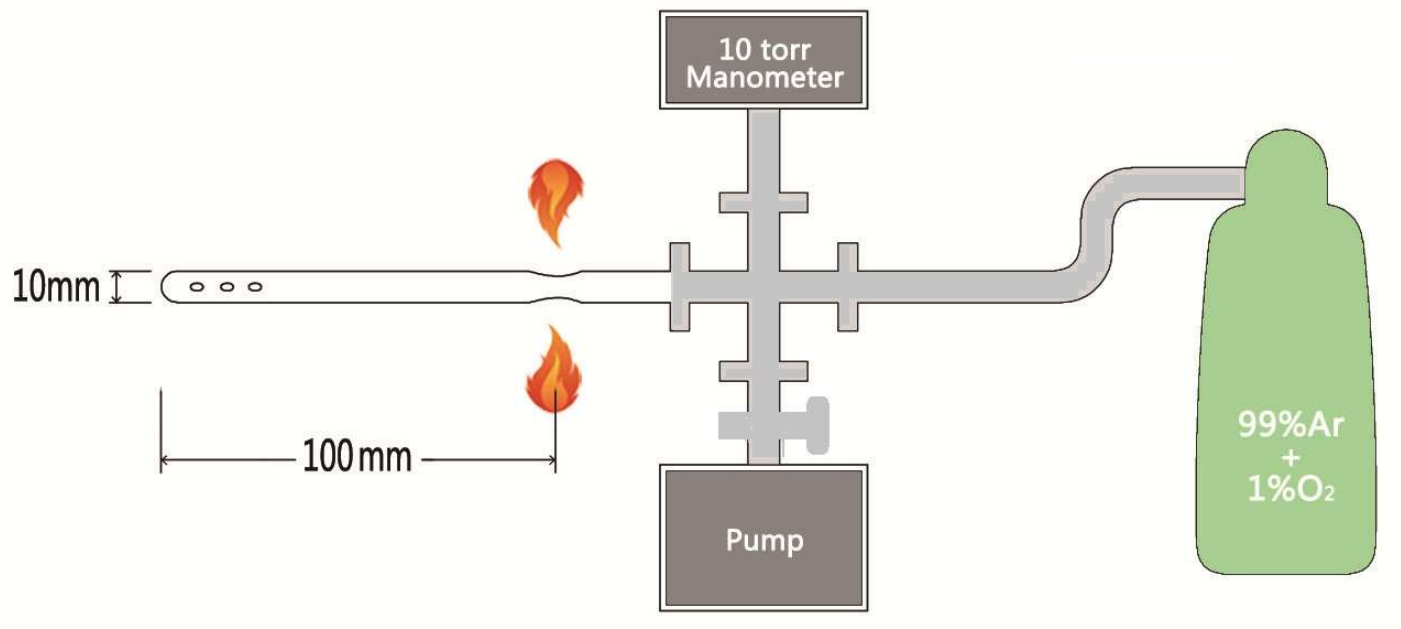}\\
	\caption{Schematic drawing of the experiment system for sealing the crystal in quartz tube with controlled amount of O$_2$.}\label{}
\end{figure} 

For the O$_2$ annealing, as-grown single crystals were cut and cleaved into thin slices with dimensions about 2.0 $\times$ 1.0 $\times$ 0.05 mm$^3$, weighed and loaded into a quartz tube with $d$ $\sim$ 10 mm. The quartz tube was carefully baked and tested not to emit any gas under the same condition as the sample annealing. The quartz tube was carefully evacuated by a diffusion pump, and the pressure was detected by a diaphragm-type manometer with an accuracy better than 1 mTorr. After totally pumping out the gas, we filled Ar/O$_2$(1\%) mixed gas into the quartz tube and sealed it into the length of 100 mm. During the sealing process, the manometer monitored the pressure in the system in real-time to prevent gas leakage and control the O$_2$ pressure in the quartz tube. The annealing system is shown schematically in Fig. 2. Because of the fixed volume, the amount of O$_2$ filled into the quartz tube can be evaluated by the pressure. Then the crystals were annealed at selected temperatures (ranging from 200 $^\circ$C to 400 $^\circ$C) for different periods of time, followed by water quenching. We also confirmed that almost all O$_2$ in the quartz tube were consumed by the crystal by breaking the quartz tube in a larger quartz tube while monitoring its pressure. Furthermore, the weight increase in the crystal after annealing is equal to the mass of O$_2$ filled into the quartz tube. Our careful experiments described above guarantee that the O$_2$ was totally absorbed by the crystal \cite{SunSciRep}. 

\begin{figure}\center	
	\includegraphics[width=8.5cm]{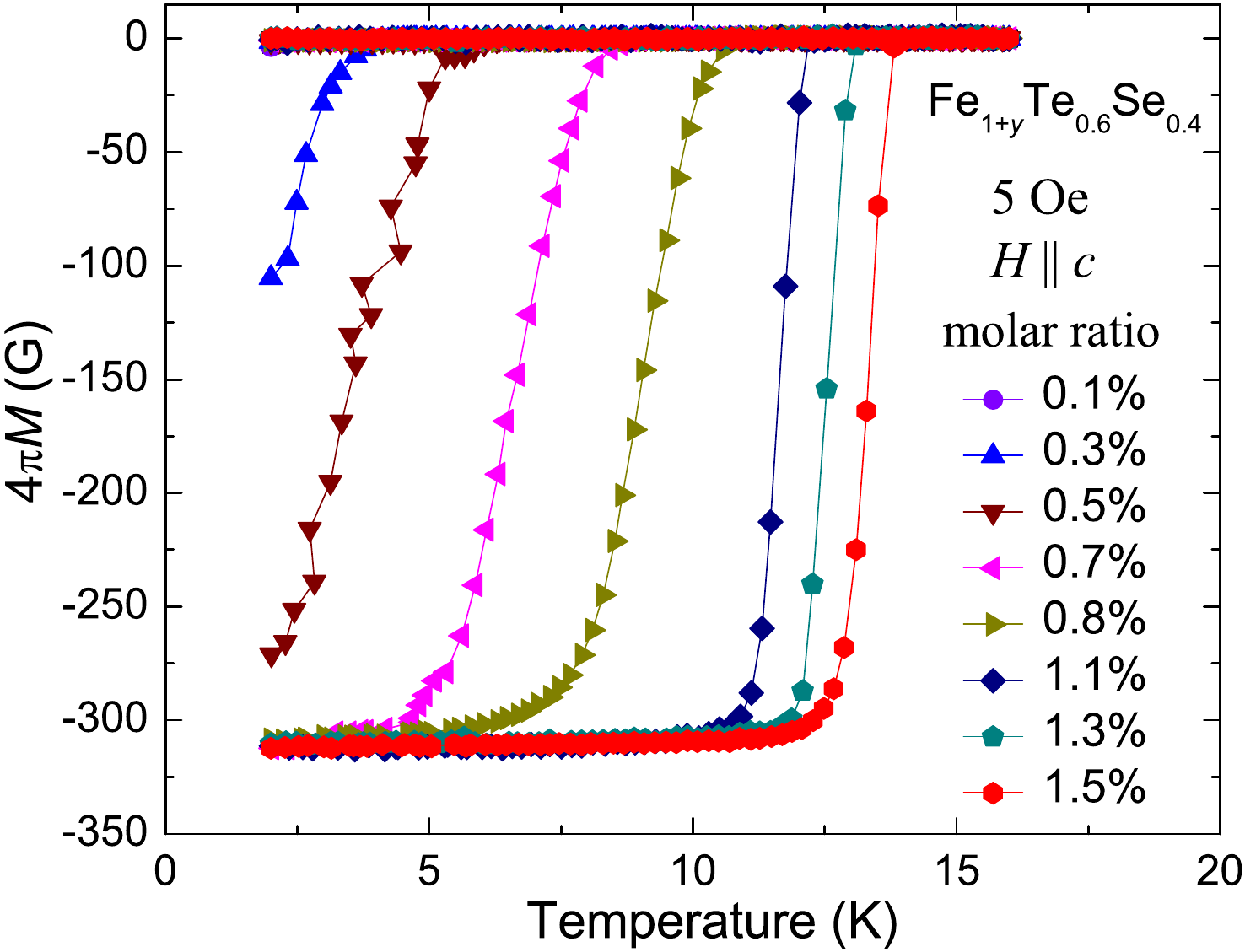}\\
	\caption{Temperature dependence of zero-field-cooled (ZFC) and field-cooled (FC) magnetization at 5 Oe for Fe$_{1+y}$Te$_{0.6}$Se$_{0.4}$ single crystal annealed at 400$^\circ$C with increasing amount of O$_2$ (molar ratio of the oxygen to the nominal Fe ranging from 0.1\% to 1.5\%) \cite{SunSciRep}.}\label{}
\end{figure}

\begin{figure}\center	
	\includegraphics[width=8.5cm]{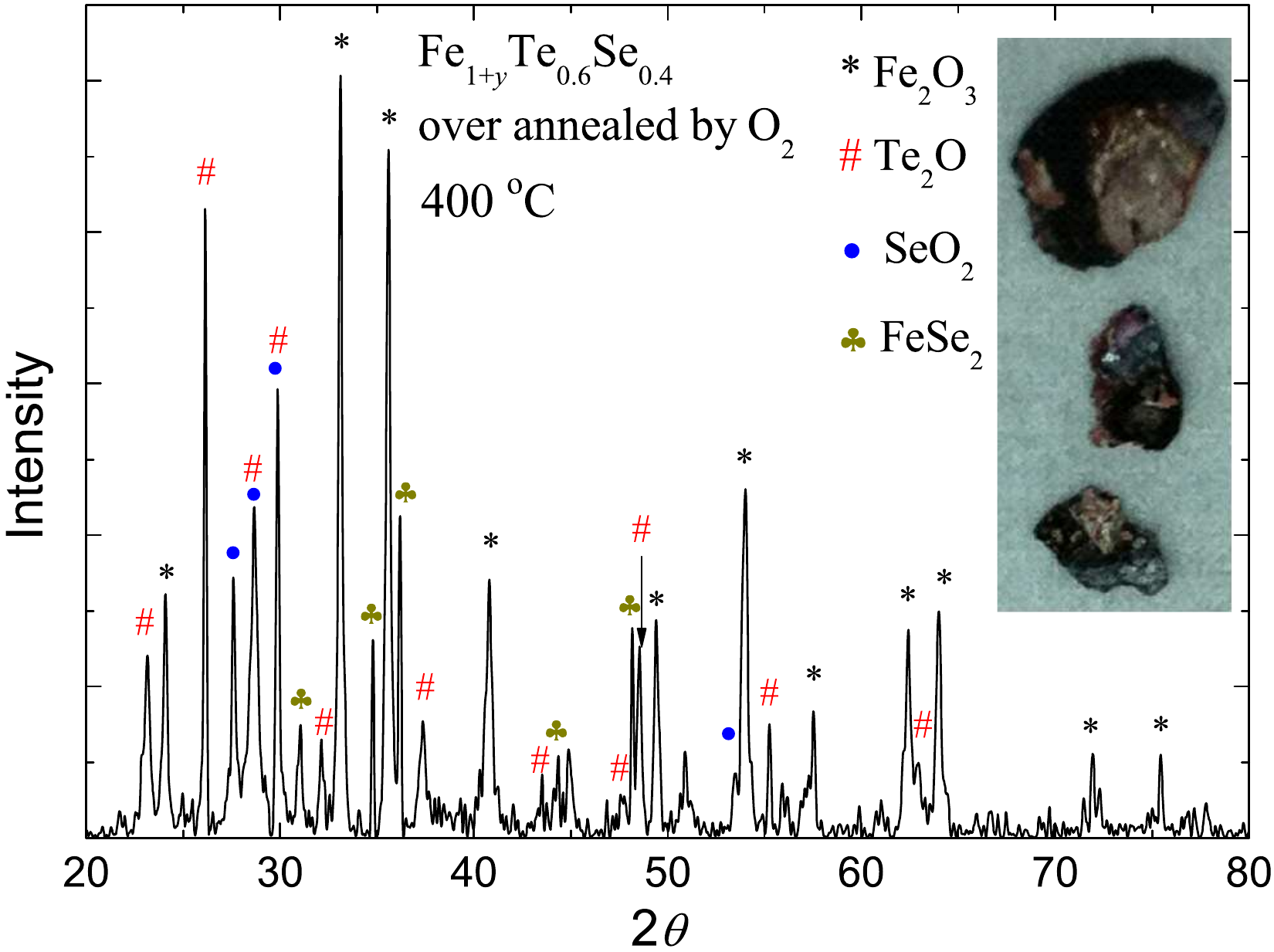}\\
	\caption{XRD pattern of the Fe$_{1+y}$Te$_{0.6}$Se$_{0.4}$ single crystal over annealed with redundant O$_2$. Inset is the photo of the over-annealed crystals. }\label{}
\end{figure} 

The O$_2$ annealing effect is demonstrated in Fig. 3, which shows temperature dependence of zero-field-cooled (ZFC) and field-cooled (FC) magnetization at 5 Oe for Fe$_{1+y}$Te$_{0.6}$Se$_{0.4}$ single crystal annealed with increasing the amount of O$_2$. The as-grown crystal usually shows no superconductivity or very weak diamagnetic signal below $\sim$ 3 K. After annealing, superconductivity emerges and $T_{\rm{c}}$ is gradually enhanced with increasing the cumulative amount of O$_2$. $T_{\rm{c}}$ reaches the maximum value over 14 K when the mole ratio of oxygen to the nominal Fe reaches $\sim$ 1.5\% ($T_{\rm{c}}$ is defined by the separating temperature for the FC and ZFC curves). On the other hand, annealing with too much O$_2$ will damage the crystal (inset of Fig. 4 shows the crystals annealed in pure O$_2$ flow for 2 days). Redundant O$_2$ reacts with the crystal, and final totally oxidizes the whole sample as indicated by the power XRD pattern of the over annealed crystal in the main panel of Fig.4, which shows no superconductivity.       

\begin{figure}\center	
	\includegraphics[width=8.5cm]{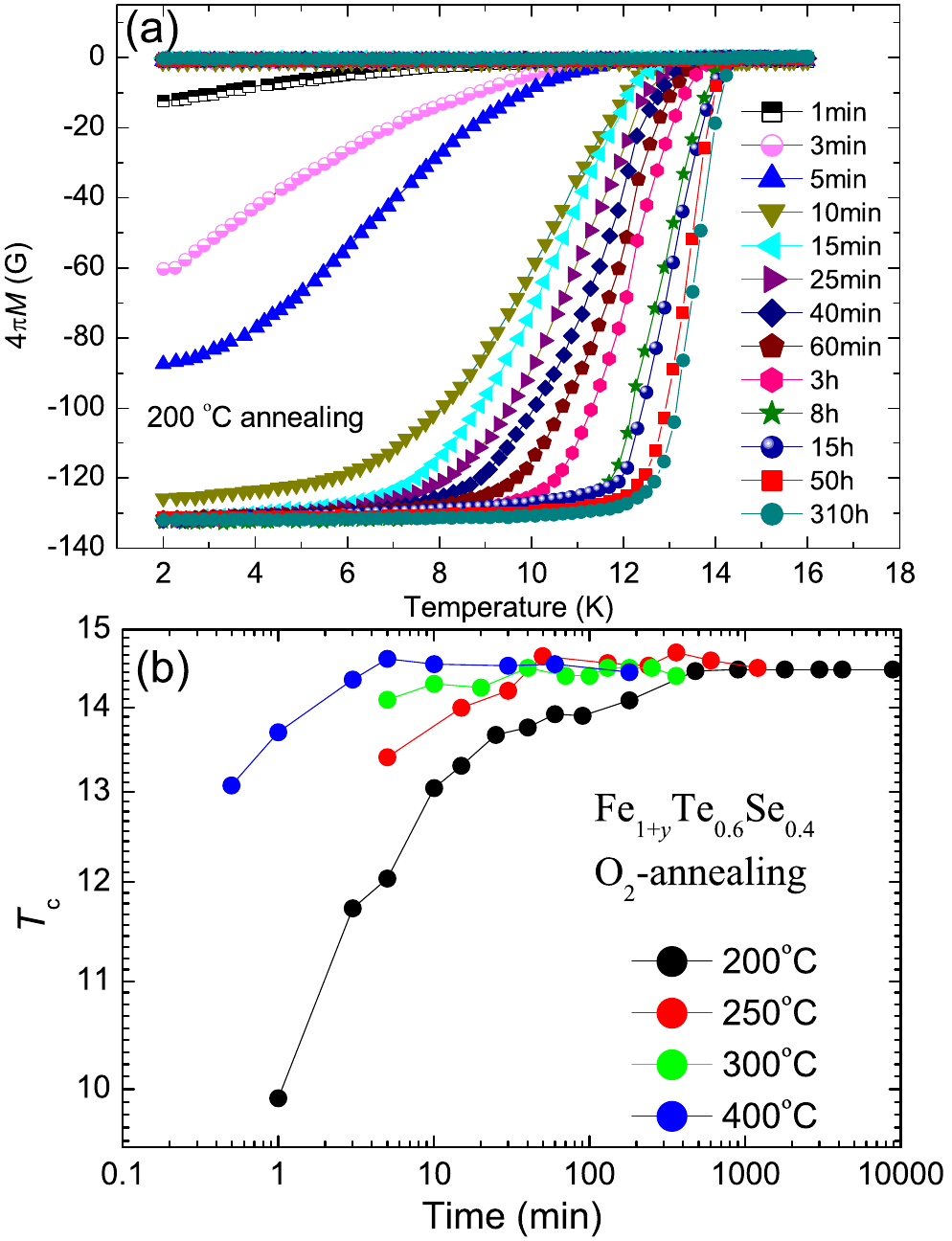}\\
	\caption{(a) Temperature dependence of magnetization at 5 Oe for Fe$_{1+y}$Te$_{0.6}$Se$_{0.4}$ single crystal annealed in O$_2$ atmosphere at 200$^\circ$C with increasing time \cite{SunSciRep}. (b) Evolution of $T_{\rm{c}}$ with time annealed in O$_2$ at temperatures of 200$^\circ$C, 250$^\circ$C, 300$^\circ$C, and 400$^\circ$C.}\label{}
\end{figure} 

Next we study the effect of annealing temperature and time. In this experiment, the crystals were annealed with the optimal amount of O$_2$ (ratio of oxygen to the nominal Fe is 1.5\%) at different temperatures with increasing time. It should be noted that the total amount of O$_2$ reacted with the crystal is larger than 1.5\% since we fix the amount of O$_2$ in all the annealing processes.  A typical result of annealing at 200$^\circ$C is shown in Fig. 5(a). With increasing annealing time, $T_{\rm{c}}$ gradually shifts to higher temperatures together with the sharpening of the transition width and the increase of diamagnetic signal. The maximum $T_{\rm{c}}$ is $\sim$ 14.3 K similar to that annealed at 400$^\circ$C (see Fig. 3). Fig. 5(b) shows the evolution of $T_{\rm{c}}$ with time annealed in O$_2$ at temperatures of 200$^\circ$C, 250$^\circ$C, 300$^\circ$C, and 400$^\circ$C. $T_{\rm{c}}$ is enhanced quicker to the maximum value when the crystal is annealed at higher temperatures. Particularly, $T_{\rm{c}}$ reaches the optimal values only after 5 min annealing at 400$^\circ$C. We also tested annealing the crystal at higher temperatures, like 500$^\circ$C. In this case, the crystal was damaged, and superconductivity was not induced. We also tested that vacuum, N$_2$, and Ar gas annealing, giving no effect to the excess Fe by using the same high vacuum sealing system. other than the O$_2$ gas, H$_2$ \cite{Friederichs_Hannealing} and F$_2$ \cite{LixiongPC} (annealed with the powers of CaF$_2$ or SmF$_3$) were reported effective to remove the excess Fe of Fe$_{1+y}$Te$_{1-x}$Se$_x$. 

\begin{figure}\center	
	\includegraphics[width=8.5cm]{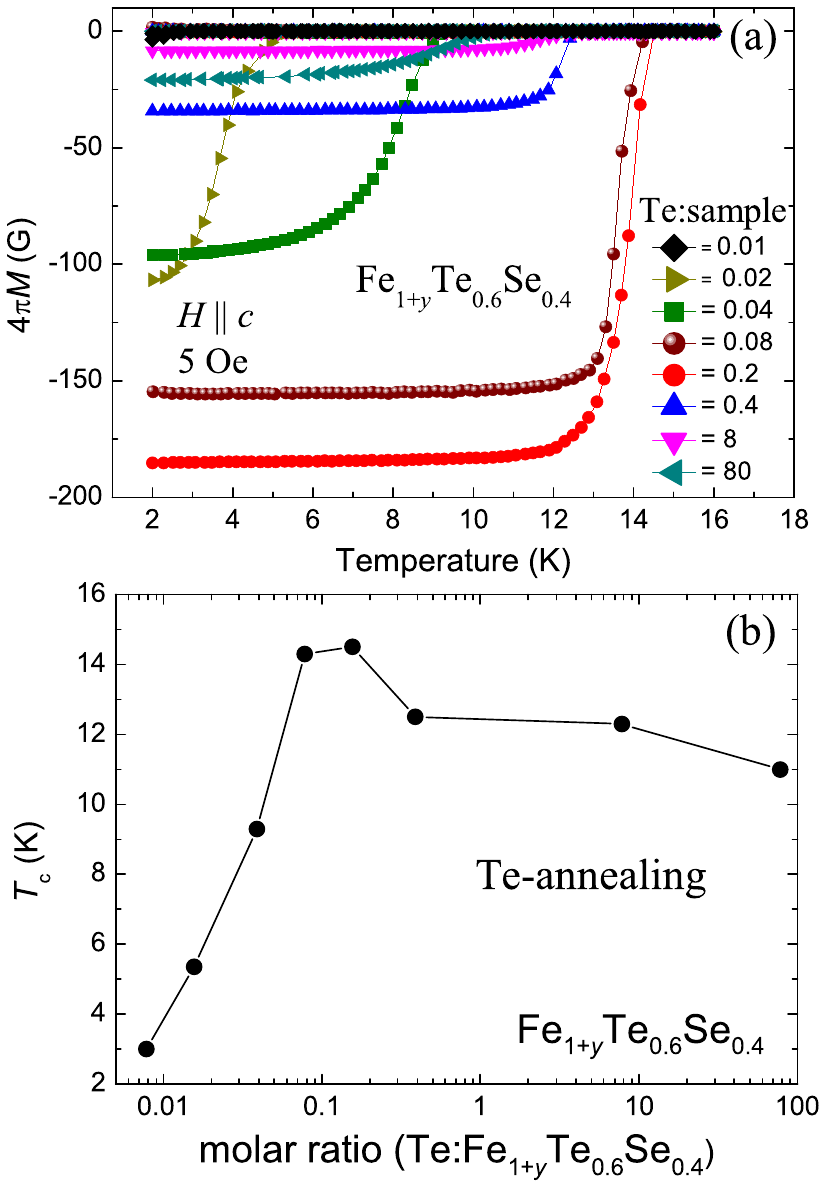}\\
	\caption{(a) Temperature dependence of zero-field-cooled (ZFC) and field-cooled (FC) magnetization at 5 Oe for Fe$_{1+y}$Te$_{0.6}$Se$_{0.4}$ single crystals annealed at 400 $^{\circ}$C with increasing amount of Te vapor (molar ratio of Te to the sample ranging from 0.01 to 80) \cite{Sunjpsj}. (b) $T_{\rm{c}}$ as functions of molar ratio of Te to Fe$_{1+y}$Te$_{0.6}$Se$_{0.4}$ single crystal annealed at 400$^{\circ}$C \cite{Sunjpsj}.}\label{}
\end{figure} 

\begin{figure}\center	
	\includegraphics[width=15cm]{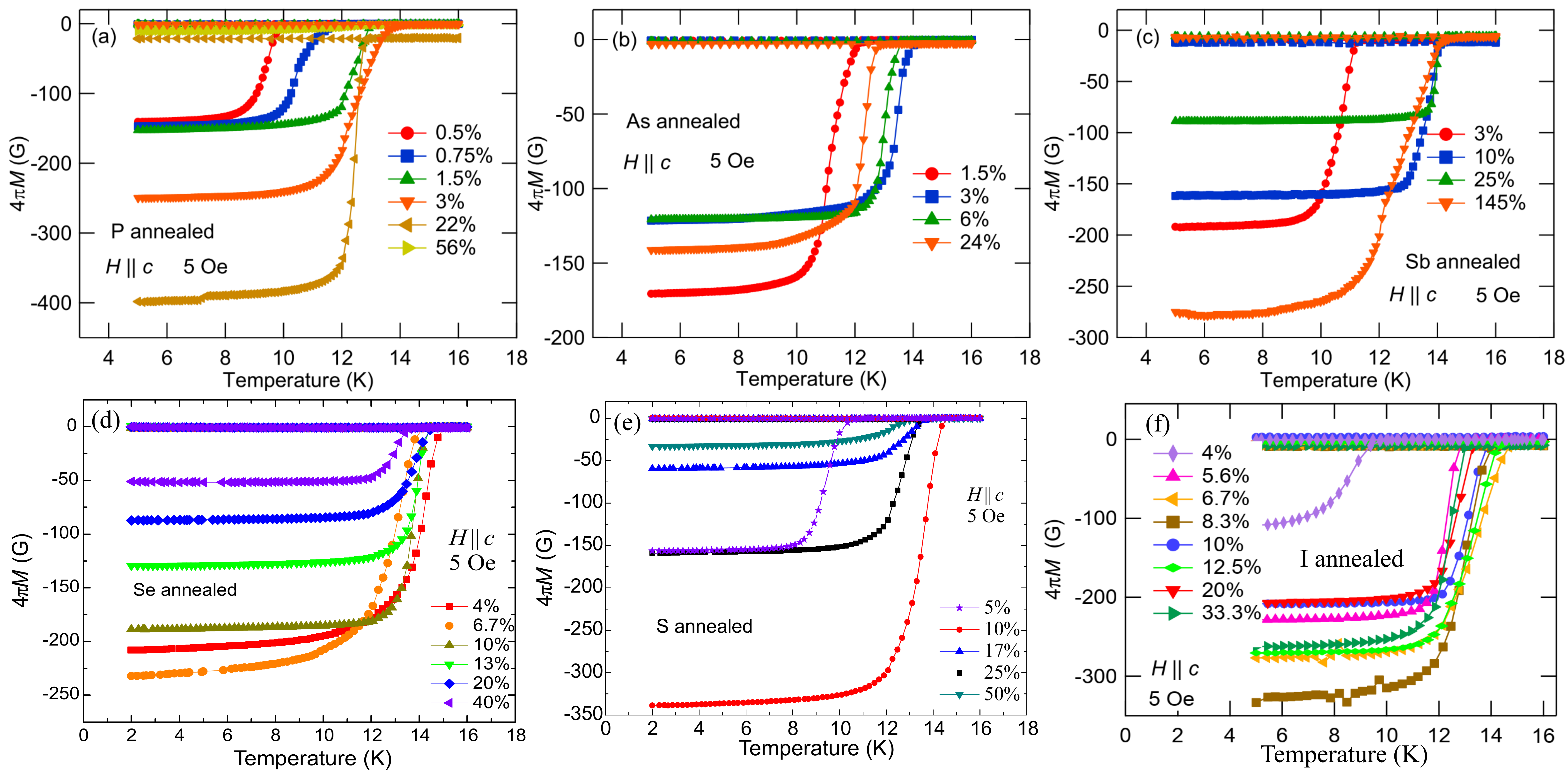}\\
	\caption{Temperature dependence of zero-field-cooled (ZFC) and field-cooled (FC) magnetization at 5 Oe for Fe$_{1+y}$Te$_{0.6}$Se$_{0.4}$ single crystals annealed at 400 $^{\circ}$C in the atmospheres of (a) P, (b) As, (c) Sb, (d) Se, (e) S, and (f) I \cite{Sunjpsj,SunJPSJshort,YamadaJPSJPanneal,ChenI2annealJPSJ}. The numbers in the legend are the molar ratio of the atmosphere elements to the sample.}\label{}
\end{figure} 

The annealing atmospheres are not restricted to the gas at room temperature and ambient pressure. Elements with vapor pressure larger than the vacuum level of the annealing system can be used to supply the annealing atmospheres. By now, elements such as Te, Se, S, P, As, I, and Sb have been confirmed to provide effective atmospheres to remove excess Fe \cite{Sunjpsj,KoshikaTeannealJPSJ,SunJPSJshort,YamadaJPSJPanneal,RodriguezIanneal,ChenI2annealJPSJ,ZhouweiAsanneal}. In our experiments, the same system presented in Fig. 2 was used to vacuum sealing the crystals. Different from the O$_2$ annealing, required amount of elements (Te, Se, S, P, As, I, or Sb) was weighted and vacuum sealed together with the single crystal. Then they were annealed at selected temperatures, such as 400$^\circ$C for 20 h, followed by water quenching. Owing to the high vapor pressure of iodine at room temperature, it would be vaporized during the evacuation. Therefore, the bottom part of the quartz tube containing the iodine and crystals was immersed into a mixture of ice and salt (cooling down to $\sim$ 258 K) to prevent the vaporizing of iodine \cite{ChenI2annealJPSJ}. Fig. 6(a) shows the temperature dependence of zero-field-cooled (ZFC) and field-cooled (FC) magnetization at 5 Oe for Fe$_{1+y}$Te$_{0.6}$Se$_{0.4}$ single crystal annealed with increasing amount of Te (molar ratio of the Te to the sample ranging from 1 :100 to 80 : 1). The evolution of $T_{\rm{c}}$ as a function of the amount of Te is summarized in Fig. 6(b). It is clear that $T_{\rm{c}}$ is gradually enhanced by annealing with increasing amount of Te, and reach the optimal values of $\sim$ 14.3 K when the molar ratio of Te to the sample is $\sim$ 0.1. After that, $T_{\rm{c}}$ decreases with increasing the amount of Te. Similar results were also observed in P, As, Sb, Se, S, and I annealings as shown in Figs. 7(a)-(f), respectively \cite{SunJPSJshort,YamadaJPSJPanneal,ChenI2annealJPSJ}. The well-annealed crystals all show the superconducting volume $\sim$ 100\% similar to that annealed in O$_2$ atmosphere.   

\begin{figure}\center	
	\includegraphics[width=8.5cm]{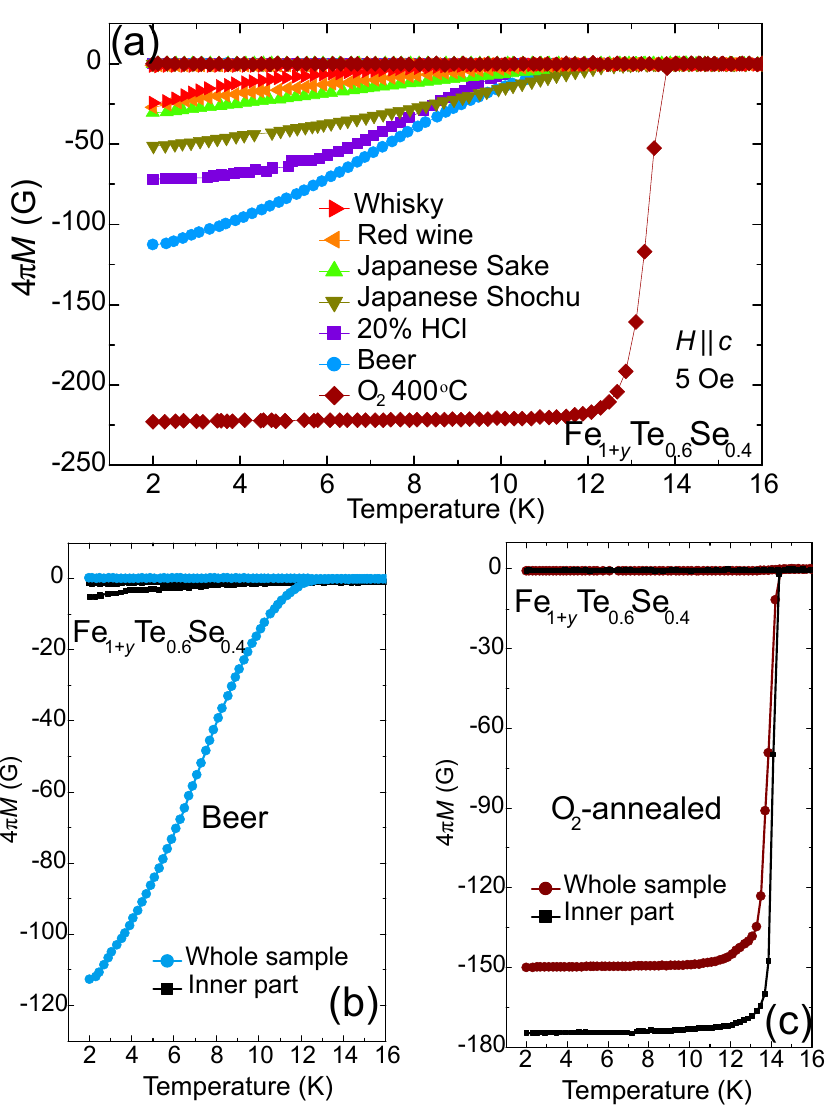}\\
	\caption{(a) Temperature dependence of zero-field-cooled (ZFC) and field-cooled (FC) magnetization at 5 Oe for Fe$_{1+y}$Te$_{0.6}$Se$_{0.4}$ single crystals immersed into alcoholic beverages (beer, red wine, Japanese sake, shochu, whisky), and 20\% HCl, together with the result of O$_2$-annealing for comparison \cite{SunSUST}. Temperature dependence of magnetization of the whole and the inner part of (b) beer treated, and (c) O$_2$-annealed Fe$_{1+y}$Te$_{0.6}$Se$_{0.4}$ single crystals \cite{SunSUST}.}\label{}
\end{figure}

The removal of excess Fe can be also realized by immersing the crystals into acids \cite{HuN2HNO3annealSUST,SunSUST} or even alcoholic beverages \cite{SunSUST}. The alcoholic beverages effect on the excees Fe was first reported by Deguchi $et$ $al$ \cite{DeguchiAlcoholicSUST} on the polycrystalline FeTe$_{1-x}$S$_x$, which was followed by similar effect in FeTe$_{1-x}$Se$_x$ single crystals \cite{SunSUST}. In these experiments, several pieces of as-grown singe crystals were put into glass bottles (10 ml) filled with 20\% hydrochloric acid HCl, beer (Asahi Breweries Ltd), red wine (Asahi Breweries Ltd), Japanese sake (Hakutsure Sake Brewing, Co. Ltd), shochu (Iwagawa Jozo Co. Ltd), or whisky (Suntory Holdings Ltd). The crystals immersed in alcoholic beverages (beer, red wine, Japanese sake, shochu, and whisky) were kept at 70$^\circ$C for 40 h. The sample immersed into 20\% HCl was kept at room temperature for 100 h, because heating up to 70$^\circ$C in acid will damage the sample quickly. Superconductivity could be induced in all these crystals. However, the transition width is very broad compared with the crystal annealed by O$_2$ (see Fig. 8(a)) \cite{SunSUST}. The broad transitions are due to the fact that superconductivity induced by acid and alcoholic beverages is only near the surface. After cutting off the four edges, and polishing the top and bottom surfaces of the samples immersed in acid and alcoholic beverages (more than half of the crystal was removed), both the $T_{\rm{c}}$ and the diamagnetic signal are largely reduced (see the example of the crystal immersed in beer in Fig. 8 (b)). For comparison, the inner part of the O$_2$ annealed crystal shows similar superconducting properties as the whole crystal (see Fig. 8(c)) \cite{SunSUST}. 

\begin{figure}\center	
	\includegraphics[width=8.5cm]{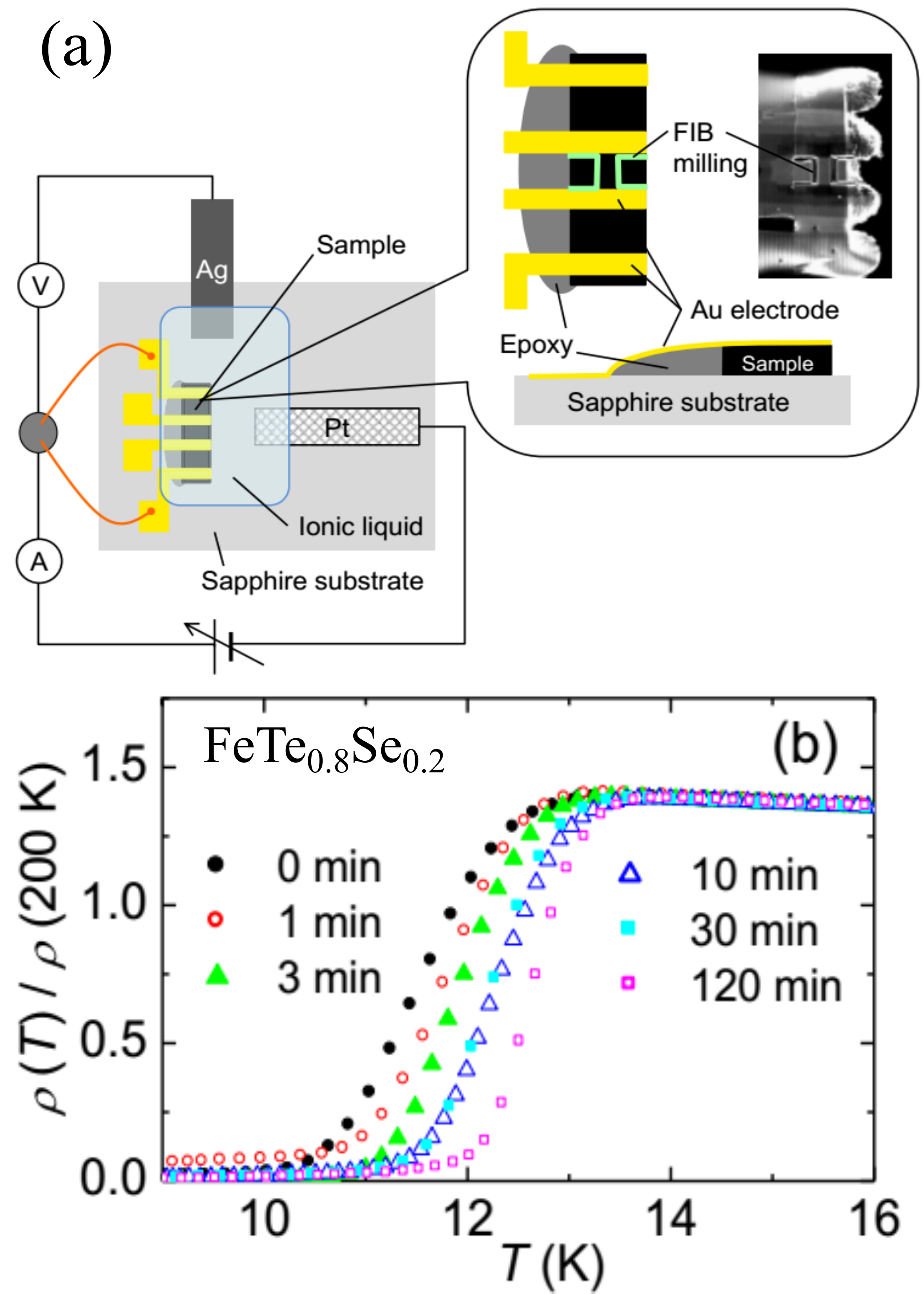}\\
	\caption{(a) Schematic of the electrochemical reaction method for a single crystal sample \cite{OkadaECJJAP}. (b) Temperature dependence of the resistivity around $T_{\rm{c}}$ for the electrochemical reaction method treated Fe$_{1+y}$Te$_{0.8}$Se$_{0.2}$ single crystal \cite{OkadaECJJAP}.}\label{}
\end{figure}

The acid and alcoholic beverage effects were further improved to be a more controllable electrochemical reaction method performed by a standard three-electrode method \cite{YamashidaSSC,YamashitaJPSJ}. A Ag plate and a Pt mesh were used as the reference and counter electrodes, respectively, while an FeTe$_{1-x}$S$_x$ sample was used as the working electrode. The ionic liquid 1-butyl-3-methylimidazolium tetrafluoroborate (BMITFB) was used as the electrolyte, and was kept at 80$^\circ$C during electrochemical reactions. Voltage increasing at a rate of 0.15 V/min up to a set value (typically 0.8 V) was applied between the reference and working electrodes and kept for a constant time. Schematics of the electrochemical reaction method is shown in Fig. 9(a) \cite{OkadaECJJAP}. The removal of excess Fe was suggested based on the increase of $T_{\rm{c}}$ (see Fig. 9(b)). The electrochemical reaction was also reported to be accelerated by fabricating the crystal into narrow bridge \cite{OkadaECJJAP}. However, there is still no bulk evidence of the superconductivity induced by electrochemical reaction method.         

\subsection{Evolution of structure, composition, and morphology with annealing}

\begin{figure}\center	
	\includegraphics[width=12cm]{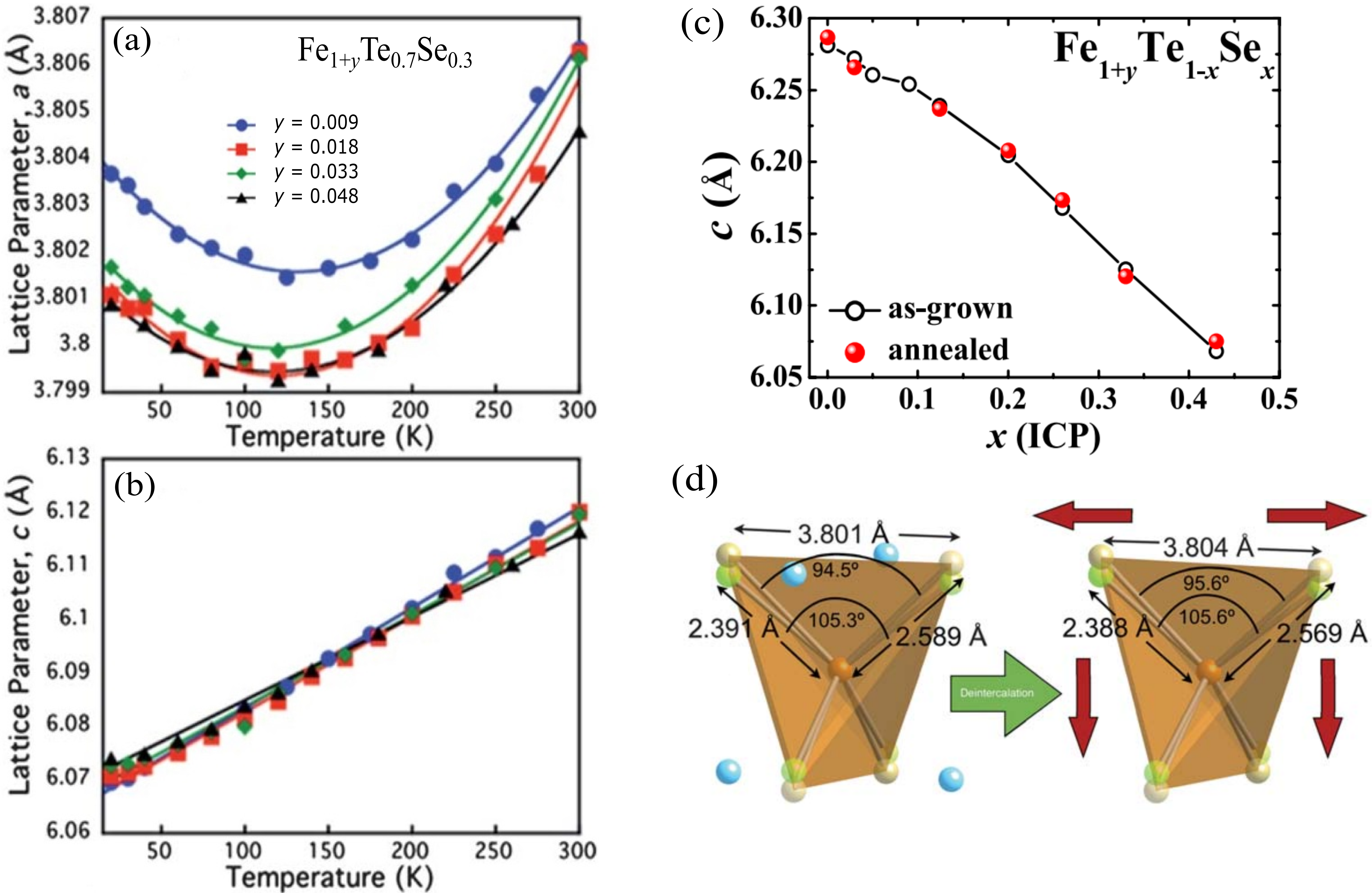}\\
	\caption{Temperature dependence of (a) the lattice constant $a$ and (b) $c$ for Fe$_{1+y}$Te$_{0.7}$Se$_{0.3}$ with $y$= 0.009, 0.018, 0.033, and 0.048 \cite{RodriguezIanneal}. (c) Lattice constant $c$ for Fe$_{1+y}$Te$_{1-x}$Se$_x$ (0 $\leq x \leq$0.43) before and after removing the excess Fe \cite{SunPDSciRep}. (d) Comparison of the structures at 7 K of Fe$_{1+y}$Te$_{0.7}$Se$_{0.3}$ before and after removal of the excess Fe, showing the deintercalation process increases the Fe–Fe distances while slightly squashing the Fe tetrahedron in the $c$ direction \cite{RodriguezIanneal}. (a), (b), and (d) Reprinted with permission from \cite{RodriguezIanneal}. Copyright 2011 by the Royal Society of Chemistry.}\label{}
\end{figure} 

Crystal structure change after removing the excess Fe was carefully studied by the powder neutron diffraction measurements on crystal annealed by iodine \cite{RodriguezIanneal}. The lattice constant $a$ was found to be increased after annealing (see Fig. 10(a)). The expansion of the $ab$ plane on reduction of the excess Fe reflects the attractive bonding involving the excess Fe within the van der Waals layers. On the other hand, the lattice constant $c$ keeps almost unchanged after annealing (see Fig. 10(b)), which was further confirmed in Fe$_{1+y}$Te$_{1-x}$Se$_x$ with $x$ ranging from 0 to 0.43 (Fig. 10 (c)). A graphical comparison of the annealing effect on the crystal structure was shown in Fig. 10(d). Similar structural change after annealing was also reported based on the XRD refinement \cite{BendeleXRDstructurePhysRevB.82.212504}.        

The amount of excess Fe usually depends on the growth condition, which was reported ranging from several percents to $\sim$ 15\% depending on groups. Even in the same batch of crystals, the amount of excess Fe is not constant. It was reported that the crystals with nice mirror-like surface usually contains more excess Fe than the crystals with textured surface. However, the textured surface is associated with slight misorientation of grains due to strain effects that develop on cooling in the constrained cross section of the quartz tube \cite{Wenjinshengreview}. To avoid the uncontrollable influence from the misorientation of grains, we prefer to use the well-orientated crystals with mirror-like surface. In our as-grown single crystals, the amount of excess Fe is $\sim$ 14\% as analyzed by inductively-coupled plasma (ICP) atomic emission spectroscopy \cite{SunSciRep}. Here, we want to emphasize that the surface composition analyses techniques like the energy dispersive X-ray spectroscopy (EDX) and electron probe microanalyzer (EPMA) usually underestimated the amount of excess Fe. It is due to the inhomogeneous distribution of excess Fe in the crystal, and the the excess Fe seems more stably located in the bulk of the sample than the surface. The EDX analyses on the same as-grown crystals for ICP gave us a value of excess Fe only about 1 - 3\%. Similar observation was also claimed by other groups \cite{RodriguezIanneal,VIENNOIS2010769}.           

\begin{figure}\center	
	\includegraphics[width=8.5cm]{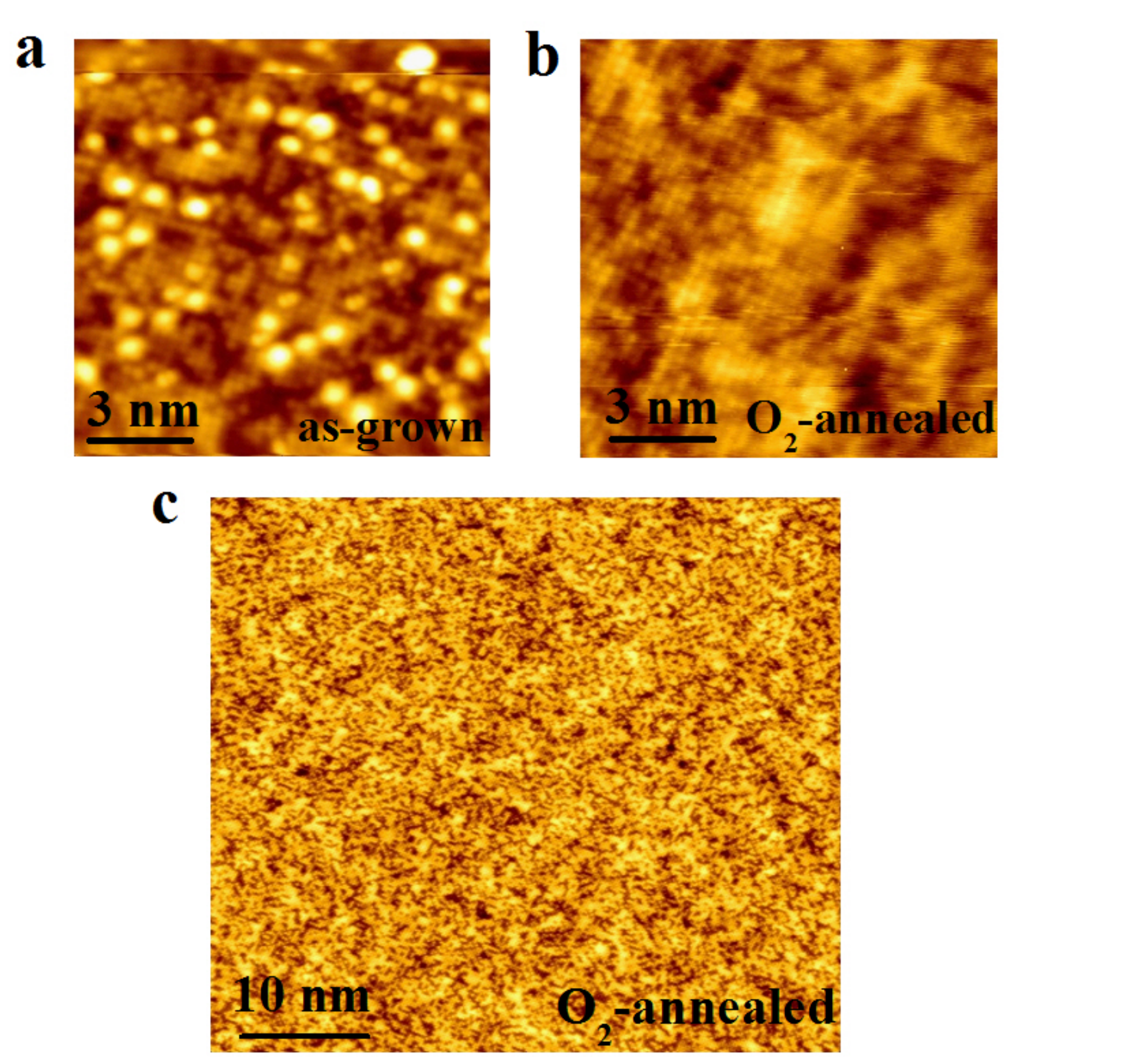}\\
	\caption{STM images for (a) as-grown, (b)–(c) O$_2$-annealed Fe$_{1+y}$Te$_{0.6}$Se$_{0.4}$ single crystal. The bright spots in (a) correspond to the excess Fe, which disappear in the optimally-annealed crystal \cite{SunSciRep}.}\label{}
\end{figure} 

The accurate determination of the amount of excess Fe in the annealed crystal is even more difficult, although many reports claimed that the excess Fe was reduced after annealing. Even though the excess Fe is removed after annealing, it should still remain in the crystal, mainly near the surface, in some form of oxides (This part will be discussed later). Thus, traditional compositional analysis methods like ICP, EDX, and EPMA can not reliably estimate the amount of excess Fe. Actually our ICP analyses on the O$_2$-annealed crystal showed that the amount of excess Fe is not zero. To precisely determine the change in the number of excess Fe, we employed the scanning tunneling microscopy (STM) measurement, which has atomic resolution. The excess Fe occupies the interstitial site in the Te/Se layer, and the previous report proved that the cleaved Fe$_{1+y}$Te$_{1-x}$Se$_x$ single crystal possesses only the termination layer of Te/Se \cite{MasseeSTMPRB}, which guarantees that the STM can directly observe the excess Fe in Te/Se layer without the influence of neighboring Fe layers. Shown in Fig. 11(a) is the STM image for the as grown Fe$_{1+y}$Te$_{0.6}$Se$_{0.4}$ single crystal. There are several bright spots in the image, which represent the excess Fe according to the previous STM analysis \cite{HanaguriScience474}. After optimal annealing, the bright spots, i.e. the excess Fe, disappeared in STM images as shown in Fig. 11(b) and a larger region in Fig. 11(c). Actually, STM observations were performed by two separate groups on different pieces of annealed crystals, and searched several different regions. Almost no bright spots can be found, which directly proves that the excess Fe was totally removed by annealing \cite{MachdiaarXiv,SunSciRep}.  

\begin{figure}\center	
	\includegraphics[width=13cm]{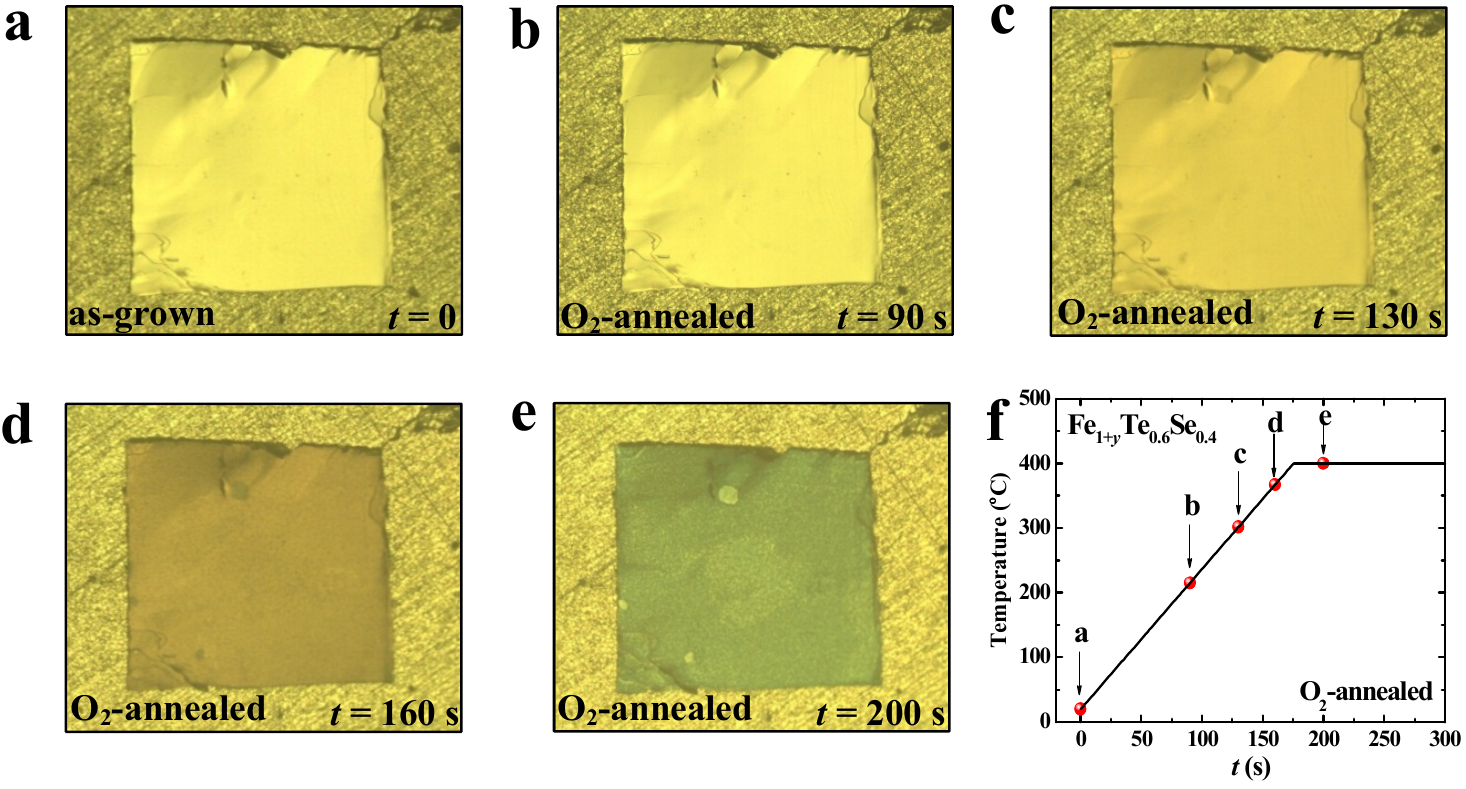}\\
	\caption{Bright view of the surface colors for the crystals (a) before and after annealed for (b) 90 s($\sim$ 200$^{\circ}$C), (c) 130 s($\sim$ 300$^{\circ}$C), (d) 160 s($\sim$ 370$^{\circ}$C), (e) 200 s($\sim$ 400$^{\circ}$C). (f) is the temperature control sequence during annealing \cite{SunSciRep}.}\label{}
\end{figure} 

Surface color change after annealing was reported in several publications \cite{DongPRB,KomiyavacumnannealJPSJ.82.064710}. To carefully observe the surface color change, we annealed the crystals with O$_2$ in a transparent furnace. The sample was quickly heated up to 400$^{\circ}$C at a rate of 130$^{\circ}$C/min as shown in Fig. 12(f). During the annealing process, a CCD camera was used to acquire the images of the sample surface. Typical images of the crystal before and after annealed for 90 s($\sim$ 200$^{\circ}$C), 130 s($\sim$ 300$^{\circ}$C), 160 s($\sim$ 370$^{\circ}$C), 200 s($\sim$ 400$^{\circ}$C) are shown in Figs. 12(a-e), respectively. Obviously, the crystal surface first changes color to brown, then turns to blue, which indicates some compounds were formed on the surface during the annealing. Here, we should point out that the color change just happens on the sample surface. After cleaving the surface layers, inner part of the crystal still keeps the mirror-like surface similar to the as-grown single crystal. The surface color change was also observed in crystals annealed in other atmospheres like the Te, Se, and S. On the other hand, the crystals annealed in vacuum did not show color change, which confirms that the color change directly relates to the annealing effect.

\begin{figure}\center	
	\includegraphics[width=8.5cm]{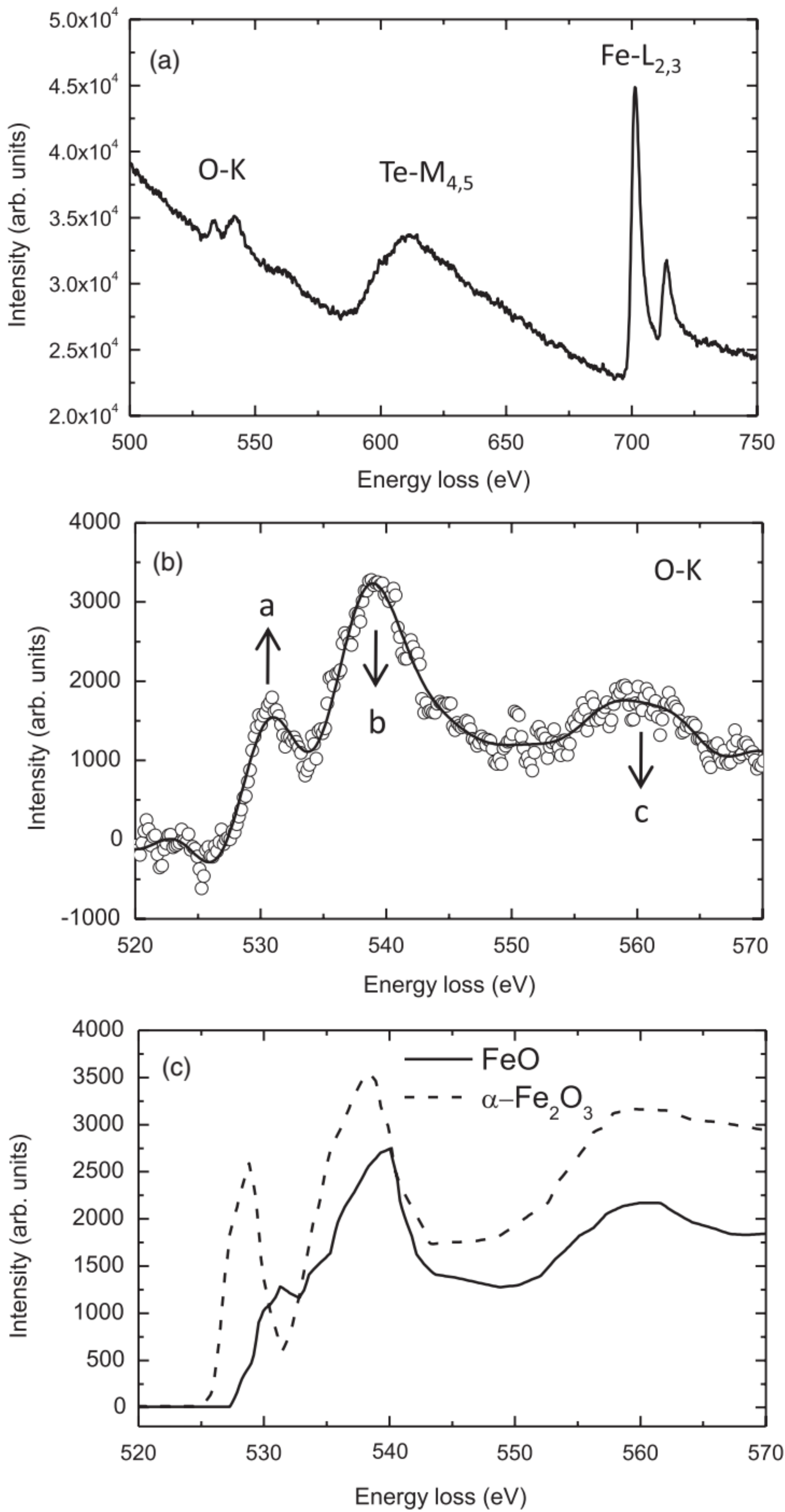}\\
	\caption{(a) EELS spectrum of the O$_2$-annealed Fe$_{1+y}$Te$_{0.5}$Se$_{0.5}$ single crystals with O-K, Te-M$_{4,5}$, and Fe-L$_{2,3}$ edges labeled. (b) The background subtracted experimental data of the O-K edge are shown as the open symbols. The solid curve, obtain by smoothing the experimental data, shows three peaks (a)–(c) indicated by the arrows, respectively. The prepeak (a) indicates hybridization of Fe 3d and O 2p states. (c) Fine structure of FeO and $\alpha$-Fe$_2$O$_3$. Reprinted with permission from \cite{HuEELSO2PhysRevB.85.064504}. Copyright 2012 by the American Physical Society.}\label{}
\end{figure} 

In the O$_2$-annealed single crystals, oxygen was reported to be detected by using electron energy loss spectroscopy (EELS) \cite{HuEELSO2PhysRevB.85.064504}, and the EPMA \cite{KomiyavacumnannealJPSJ.82.064710}. Fig. 13(a) shows a typical EELS spectrum with O-K, Te-M$_{4,5}$, and Fe-L$_{2,3}$ edges labeled. The peaks around 530 eV indicates the  Fe 3d level hybridized with oxygen 2p level. After subtracting the background, O-K edge is shown in Fig. 13(b). For comparison, the fine structure of O-K edges of FeO and $\alpha$-Fe$_2$O$_3$ in the same energy range are plotted in Fig. 13(c). Similar EELS spectra shown in Fig. 13(b) and (c) indicate the existence of the FeO and/or $\alpha$-Fe$_2$O$_3$ in the annealed surface layers \cite{HuEELSO2PhysRevB.85.064504}. Fig. 14 shows the EPMA results of the crystal annealed in poor vacuum (It is actually the O$_2$-annealing based on our discussion before.). The oxygen was found only on the surface layers of the annealed crystal, and the surface layers contain only the elements of Fe and O. Oxygen was not detected in the inner part of the annealed crystals after peeling off the surfaces, which proves that the oxygen is not doped into the crystal \cite{KomiyavacumnannealJPSJ.82.064710}. 

\begin{figure}\center	
	\includegraphics[width=8.5cm]{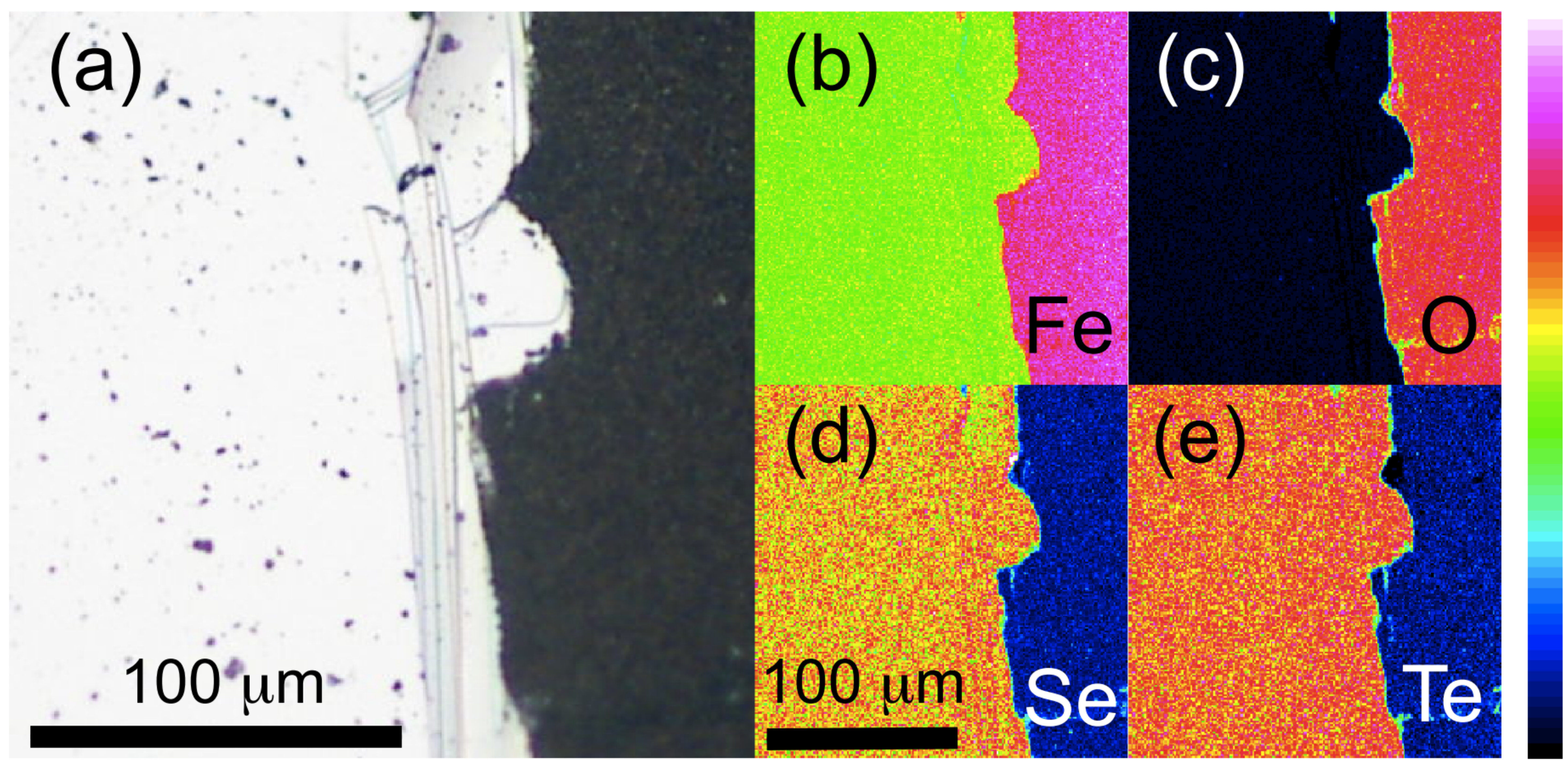}\\
	\caption{(a) Optical microscope image of the annealed crystal. (b)–(e) Elemental mapping of Fe, O, Se, and Te for the same area as (a). The left-hand side of the area is peeled to remove surface black layers, and the right-hand side is as annealed surface. On the as annealed surface, almost no Se or Te is detected. Reprinted with permission from \cite{KomiyavacumnannealJPSJ.82.064710}. Copyright 2013 by the Physical Society of Japan.}\label{}
\end{figure} 

The reduction of excess Fe in the inner part of the crystal, and the formation of the FeO$_x$ surface layers indicate that the excess Fe reacts with O$_2$ and moves to the surface of the crystal. However, it is well known that the detection of oxygen is relatively difficult for the EPMA and EELS measurements. To confirm this point, we chose the crystals annealed in As atmosphere \cite{ZhouweiAsanneal}, which is an exotic and easily detectable heavy element to Fe$_{1+y}$Te$_{1-x}$Se$_x$. Carefully measurements of the compositions of the inner and surface layers by the ICP and the EDX measurements confirmed that Fe reacts with As on the surface of the crystal and the reaction itself acts as a driving force to drag excess Fe out.

\begin{figure}\center	
	\includegraphics[width=8.5cm]{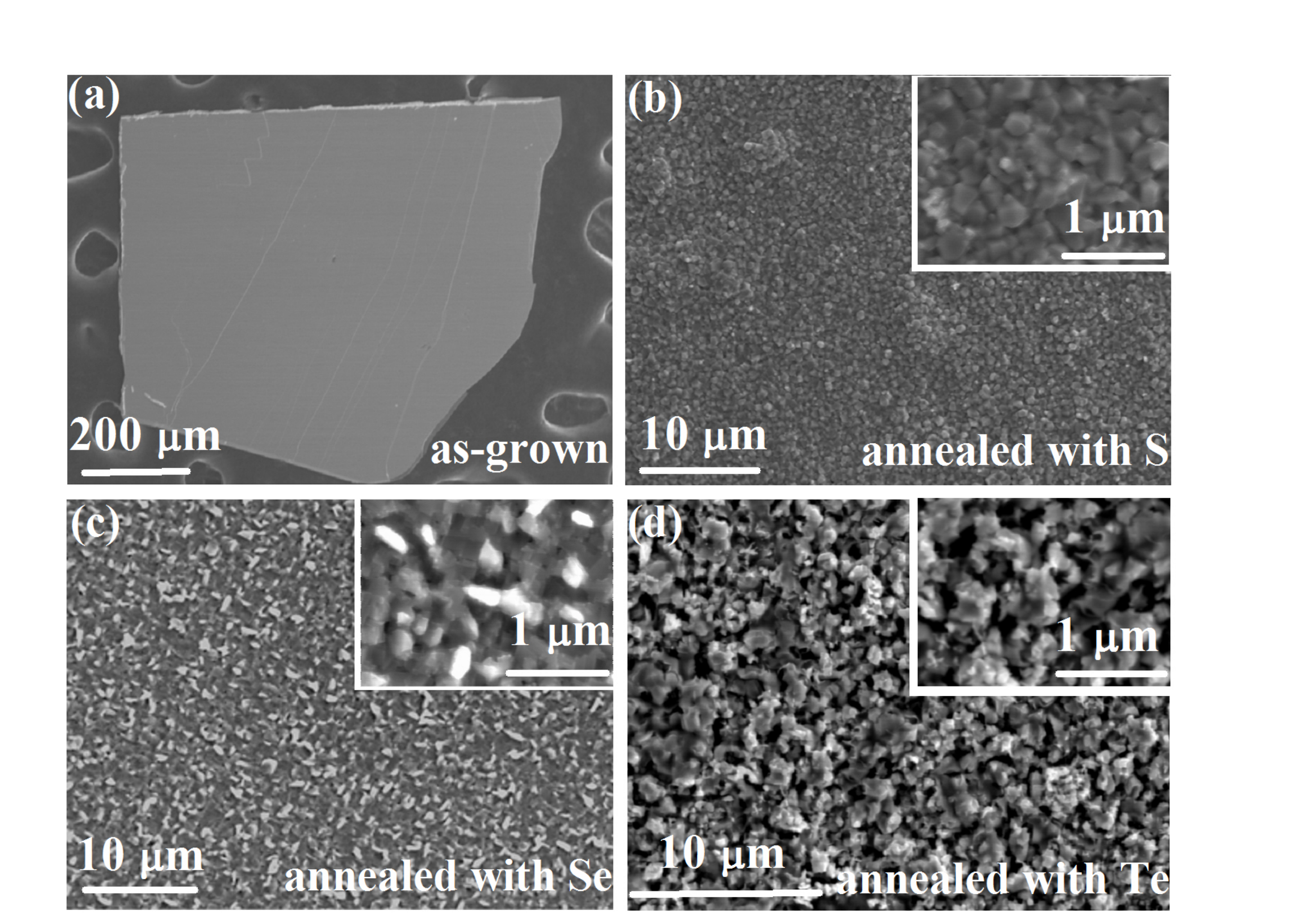}\\
	\caption{Scanning electron microscope (SEM) images for (a) as-grown, (b) S-annealed, (c) Se-annealed, and (d) Te-annealed Fe$_{1+y}$Te$_{0.6}$Se$_{0.4}$ single crystals. Insets are enlarged images \cite{SunPhyscaCTeSeS}.}\label{}
\end{figure} 

For crystals annealed in other atmospheres, the formation of surface layers was also observed. Figs. 15(b–d) shows SEM images for the surface of the Fe$_{1+y}$Te$_{0.6}$Se$_{0.4}$ single crystals annealed in S, Se, or Te atmospheres \cite{SunPhyscaCTeSeS}. Obviously, the mirror-like surface of the as-grown crystal turns into polycrystal-like. EDX measurements shows that the polycrystal-like surface for sample annealed in S/Se contains only Fe and S/Se with a molar ratio roughly 1:1, while the molar ratio is roughly 1:2 in the surface layer of Te annealed sample. X-ray diffraction (XRD) measurements show that the surface layers are most probably FeS, Fe$_7$Se$_8$, and FeTe$_2$ for crystal annealed in the atmosphere of S, Se or Te, respectively. The formation of the FeM$_x$ (M represents the elements used for providing the annealing atmospheres, such as the O, S, Se, Te, Sb, P, As) was also observed in many other publications \cite{Sunjpsj,SunJPSJshort,YamadaJPSJPanneal,ZhouweiAsanneal,SunPhyscaCTeSeS,LinWenzhiPhysRevB.91.060513}.

\subsection{Mechanism of annealing}

\begin{figure}\center	
	\includegraphics[width=8.5cm]{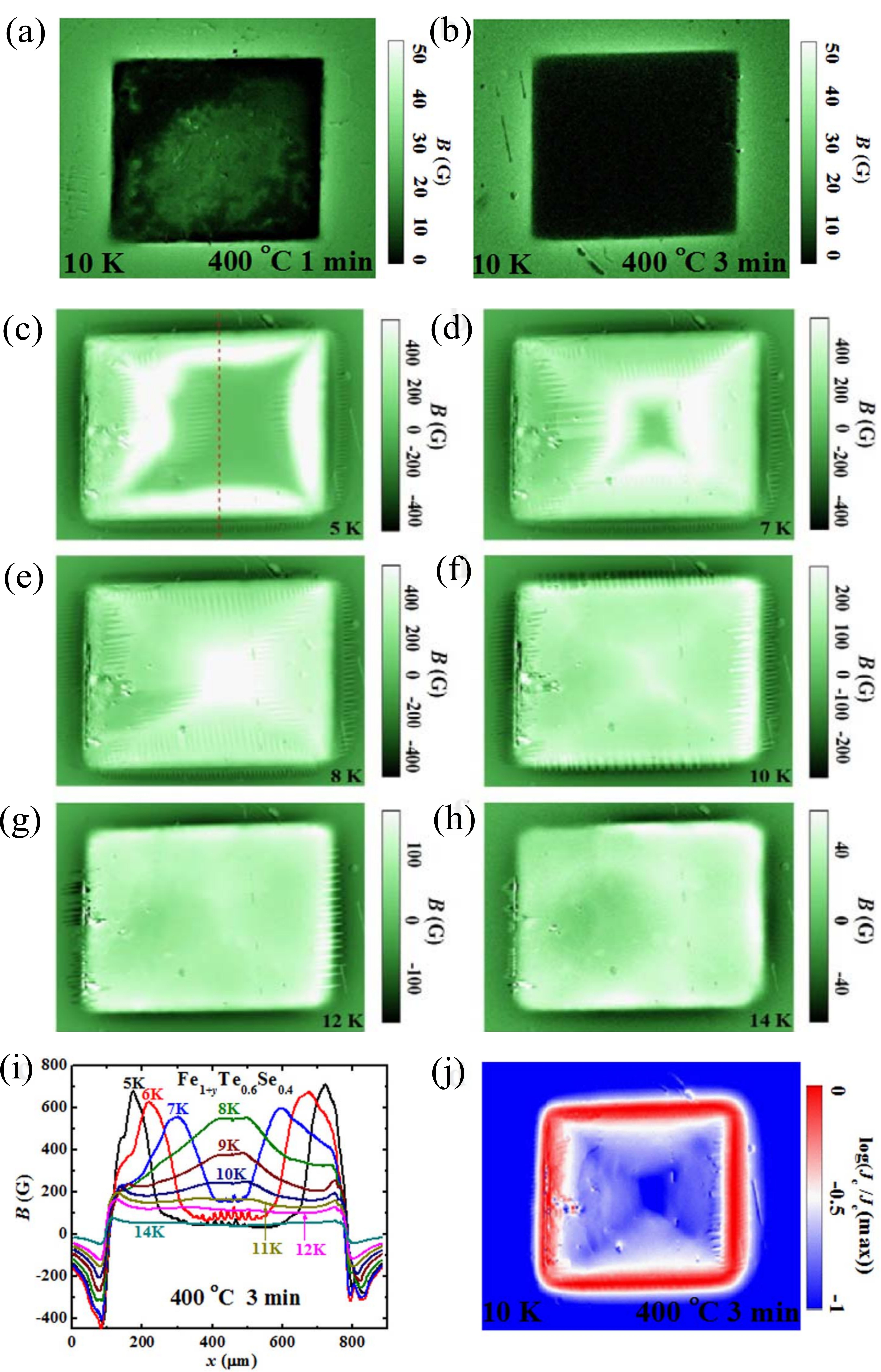}\\
	\caption{Meissner state magneto-optical (MO) images under 5 Oe at 10 K for Fe$_{1+y}$Te$_{0.6}$Se$_{0.4}$ annealed at 400$^{\circ}$C for (a) 1 and (b) 3 min, respectively. Magneto-optical images in the remanent state at (c) 5 K, (d) 7 K, (e) 8 K, (f) 10 K, (g) 12 K, and (h) 14 K for the crystal annealed for 3 mins. This state is prepared by applying 400 Oe along the $c$-axis for 1 s and removing it after zero-field cooling. (i) Local magnetic induction profiles at temperatures from 5 K to 14 K taken along the dashed lines in (c). (j) Spatial distribution of $J_{\rm{c}}$/$J_{\rm{c}}$(max) at 10 K for crystal annealed for 3 min \cite{SunSciRep}.}\label{}
\end{figure} 

In the following, we study the spatial evolution of the annealing effect by monitoring the evolution of superconductivity in the crystal using the magneto-optical (MO) images. In this experiment, a thin crystal with perfect mirror-like surface, without cracks or damages, is used.  Figs. 16(a) and (b) show the MO images of shielding of 5 Oe field after zero-field cooling at 10 K for the crystal annealed at 400$^{\circ}$C in O$_2$ atmosphere for 1 and 3 min, respectively. When the sample is annealed for 1 min, only the edge part shows Meissner state, while the central part still remains non-superconducting. When the annealing time is increased to 3 min, all parts of the sample turn into Meissner state. It indicates that the superconductivity mainly evolves from edges of the crystal to the center \cite{SunSciRep}. 

Figs. 16(c)-(h) show the remnant state MO images at temperatures ranging from 5 to 14 K for the crystal annealed for 3 min. Fig. 16(i) shows profiles of the magnetic induction at different temperatures along the dashed line in Fig. 16(c).  Both the MO images and profiles show that the magnetic field cannot totally penetrate the sample at 5 K. At higher temperature, the field penetrates deeper into the center. At 8 K, the sample is totally penetrated, and the MO image manifests a typical roof-top pattern up to 10 K. After that, from 11 to 14 K, the edge of the crystal traps higher magnetic field than the center, which indicates the $J_{\rm{c}}$ in the edge of the crystal is larger. To directly observe the distribution of $J_{\rm{c}}$ in the crystal, we converted the MO images into the current distribution with a thin-sheet approximation \cite{SunthinfilmMOSUST} and by using the fast Fourier transform calculation method. A typical image of the distribution of the modulus of $J_{\rm{c}}$ at 10 K is shown in Fig. 16(j), which manifests that the edge of the crystal maintains a larger current, and the $J_{\rm{c}}$ decays from the edge to the center. All these results testify that superconductivity mainly evolves from the edge of the crystal to the center. It should be noted, however, that we can observe a total expulsion of magnetic flux at 5 K even in the sample annealed for only 1 min. This fact means that weak superconductivity also evolves from the top and bottom surfaces \cite{SunSciRep}.        

In a short summary, based on the crystal structure, composition, and morphology change with the annealing, it is now clear that the excess Fe was removed after annealing in appropriate atmosphere. The excess Fe is somehow attracted to the surface, reacts with the atmosphere-elements, and forms the FeM$_x$ surface layers. Besides, the annealing effect mainly evolves from the edge of the crystal to the center.  
 
\section{Annealing effects on normal state properties} 

\subsection{Annealing effects on magnetism}

The magnetic properties of Fe$_{1+y}$Te$_{1-x}$Se$_x$ are unique in IBSs. As first reported by Bao $et$ $al$. \cite{BaoWeiPRL}, the magnetic wave vector in Fe$_{1+y}$Te is in the ($\delta\pi$, $\delta\pi$) direction (along the diagonal direction of the Fe-Fe square), which is different from the ($\pi$, 0) along the $a$-axis direction in Fe-pnictide and FeSe superconductors (Fig. 17) \cite{ShiliangLiPhysRevB.79.054503}. On the other hand, the magnetic order is incommensurate in Fe$_{1+y}$Te with redundant excess Fe ($y$ $\sim$ 0.14), decreases with reducing excess Fe, and becomes commensurate in crystal with $y$ $\sim$ 0.076. The theoretical calculations find that the excess Fe with valence near Fe$^+$ is strongly magnetic, which provides lacal moments that interact with the plane Fe magnetism \cite{ZhangPRB}. In detail, the excess Fe modifies the local moments’ exchange interactions via the multiorbital generalization of the long-range Ruderman-Kittel-Kasuya-Yosida (RKKY) interaction \cite{DucatmanPhysRevB.90.165123}. Neutron scattering measurement also shows that excess Fe in superconducting Fe$_{1+y}$Te$_{1-x}$Se$_x$ induces a magnetic Friedel-like oscillation and involves more than 50 neighboring Fe sites \cite{ThampyPRL}. Those results manifest that the excess Fe does not act as a separated point-like defect, but has global effect due to the strong magnetic nature. Thus, the excess Fe affects almost all the properties of Fe$_{1+y}$Te$_{1-x}$Se$_x$ from normal state to superconducting state.             

\begin{figure}\center	
	\includegraphics[width=8.5cm]{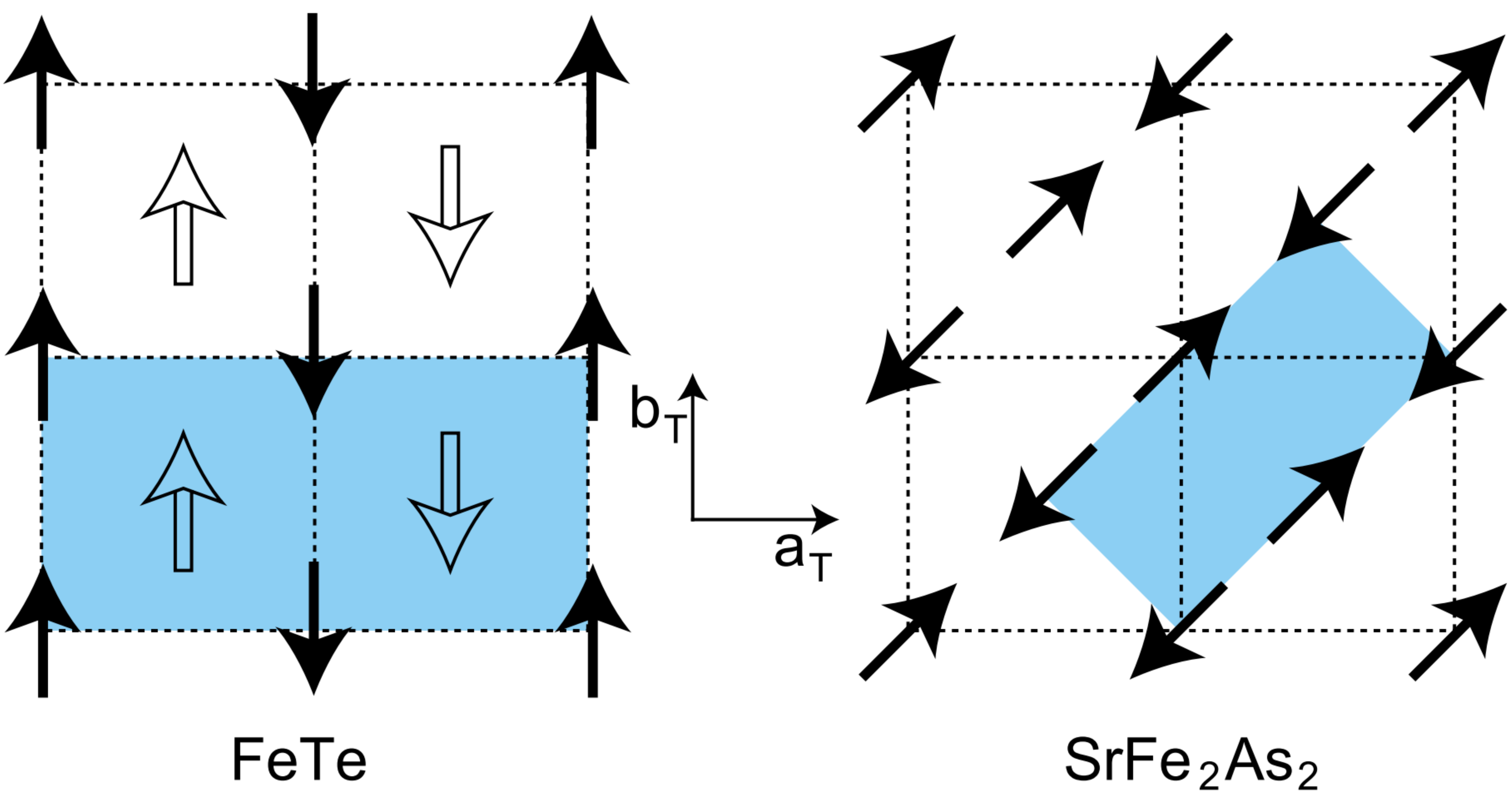}\\
	\caption{Schematic in-plane spin structure of FeTe and SrFe$_2$As$_2$. The solid arrows and hollow arrows represent two sublattices of spins. The shaded area indicates the magnetic unit cell. Reprinted with permission from \cite{ShiliangLiPhysRevB.79.054503}. Copyright 2009 by the American Physical Society.}\label{}
\end{figure} 

\begin{figure}\center	
	\includegraphics[width=8.5cm]{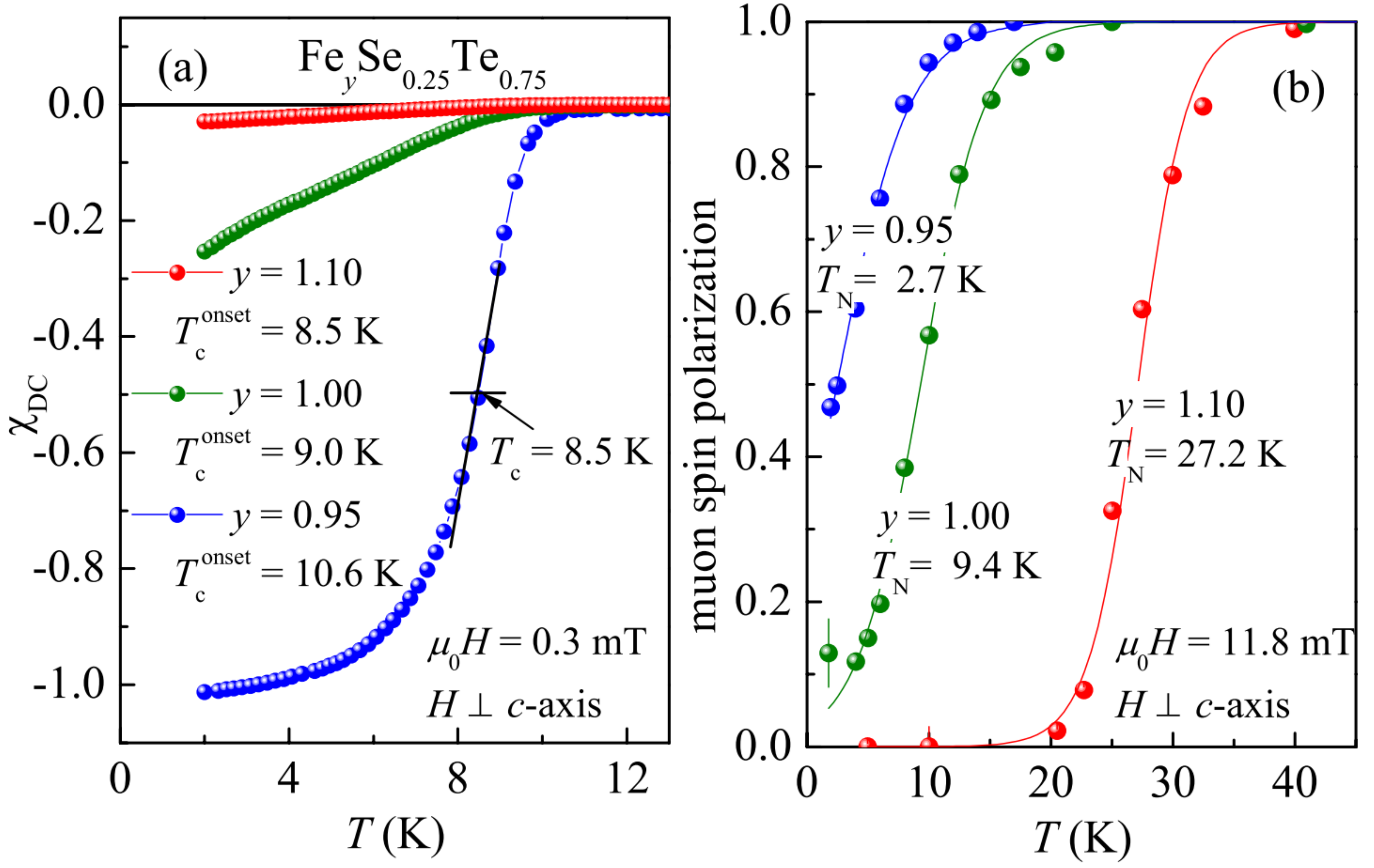}\\
	\caption{(a) Temperature dependence of the volume susceptibility $\chi_{dc}$ of representative compositions ($y$=0.95, 1.00, and 1.10 of single-crystal Fe$_y$Te$_{0.75}$Se$_{0.25}$. The onset of the superconducting transition $T_{\rm{c}}$ onset and the midpoint corresponding to $\chi_{dc}$=-0.5 are indicated. (b) Temperature dependence of the muon-spin polarization of the slow relaxing component ($P_{\rm{slow}}^{\rm{TF}}$). The magnetic transition $T_{\rm{N}}$ is determined from a fit to a Fermi-type function. Reprinted with permission from \cite{BendelePRB}. Copyright 2010 by the American Physical Society.}\label{}
\end{figure} 

The muon-spin rotation measurements on Fe$_y$Te$_{0.75}$Se$_{0.25}$ with different amount of excess Fe show that the magnetic transition temperature $T_{\rm{N}}$ is gradually suppressed by reducing excess Fe, and the $T_{\rm{c}}$ is enhanced (Fig. 18) \cite{BendelePRB}. The influence of excess Fe and the annealing effect on the magnetism of Fe$_{1+y}$Te$_{1-x}$Se$_x$ with 0 $\leq$ x $\leq$ 0.43 was systematically studied by the magnetic susceptibilities measurements under 1 T magnetic field \cite{SunPDSciRep}. Figs. 19(a) and (b) show the normalized magnetic susceptibilities for the as-grown and annealed Fe$_{1+y}$Te$_{1-x}$Se$_x$ (0 $\leq$ x $\leq$ 0.43) single crystals, respectively. The as-grown Fe$_{1+y}$Te shows a sharp transition at $\sim$ 58 K, which is due to the antiferromagnetic (AFM) transition. With Se doping, the AFM transition temperature $T_{\rm{N}}$ is gradually suppressed to lower temperatures, and becomes much broader at $x$ = 0.09. Afer that, the AFM transition disappears and is replaced by a very broad hump-like feature. Such a hump-like feature may be originated from the spin glass state according to the neutron scattering results \cite{KatayamaPhaseD}. The hump-like feature survives up to $x$ = 0.33, and is not observed for $x \geq$ 0.43. 

In the annealed crystal, the value of magnetic susceptibility does not show a systematic evolution and is irregular, which is caused by the magnetism from some Fe impurities. During the annealing, the excess Fe are removed from their original positions, and form some compounds like Fe$_2$O$_3$ or FeTe$_2$ based on the discussion above. Although those impurities are mainly formed in the surface layers, and removed by polishing, small parts may still remain inside the crystals and disturb the magnetic susceptibility value because of their strong magnetism. However, we can still obtain some important information from the data regardless of the irregularity in the absolute value. As shown clearly in Fig. 19(b), the value of $T_{\rm{N}}$ for the pure FeTe is enhanced to $\sim$ 72 K after removing the interstitial Fe. The AFM transition is only observed in crystals with $x$ = 0 and 0.03. When the Se doping level is increased over 0.05, the AFM is totally suppressed. On the other hand, the hump-like feature observed in the as-grown crystals is not witnessed after annealing. For $x >$ 0.03, the annealed crystals only show the SC transition at low temperatures.

\begin{figure}\center	
	\includegraphics[width=8.5cm]{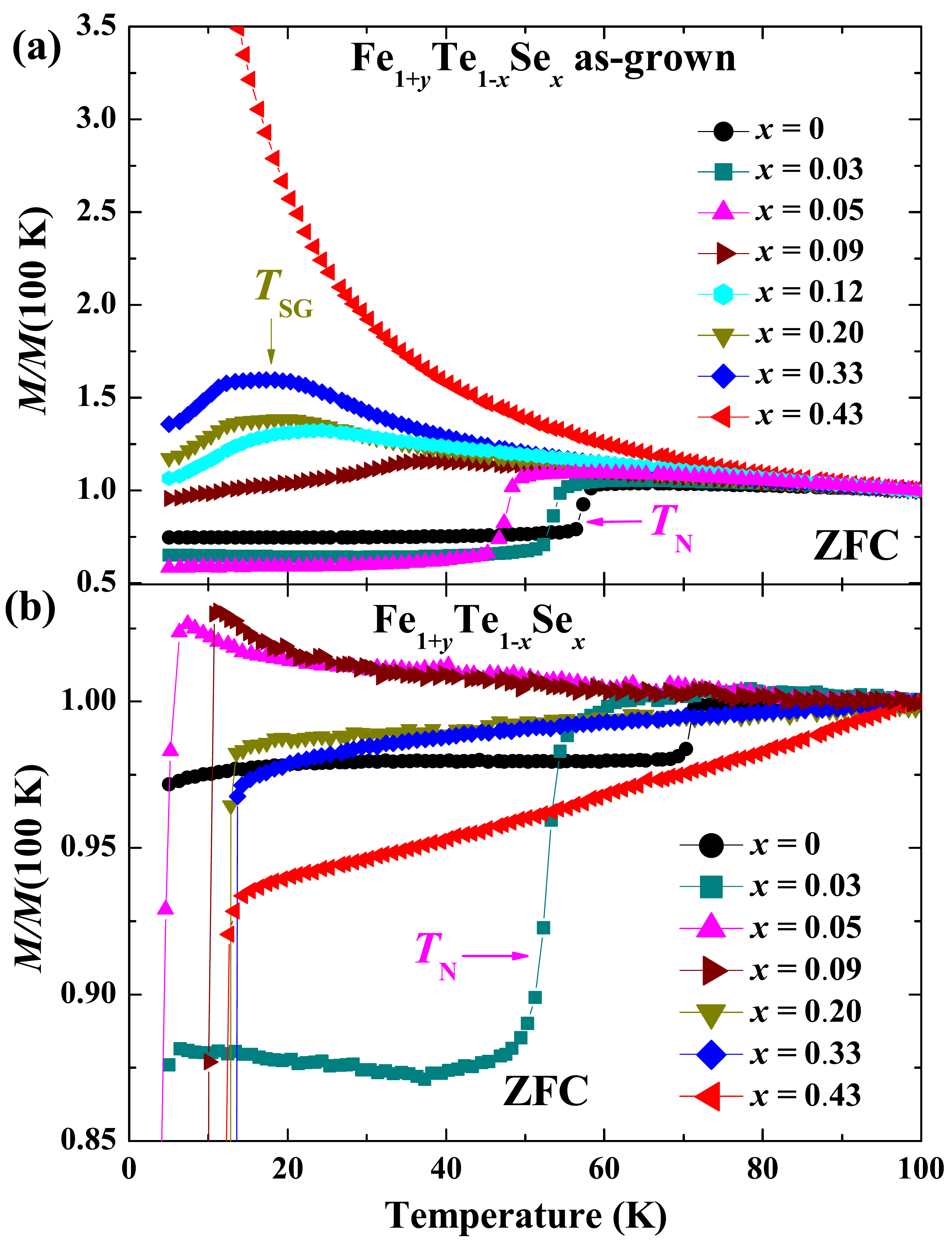}\\
	\caption{Magnetic susceptibilities measured at 1 T with $H \parallel c$ for Fe$_{1+y}$Te$_{1-x}$Se$_x$ (0 $\leq$ $x$ $\leq$ 0.43) (a) before and (b) after O$_2$-annealing \cite{SunPDSciRep}.}\label{}
\end{figure} 

\subsection{Annealing effects on transport properties}

The magnetic moment from excess Fe will localize the charge carriers affecting the normal state transport properties as reported firstly by Liu $et$ $al$. \cite{LiuPRB}. Benefit from the controllable annealing technique reported above, we studied in detail the influence of excess Fe to transport properties by measuring the resistivity, Hall effect, and the magnetoresistance (MR) in three selected crystals: the as-grown, half-annealed, and the fully-annealed ones \cite{SunPRB}. Combined ICP and STM measurements confirmed that the amount of excess Fe in the as-grown, half-annealed, and fully-annealed crystals are roughly 14\%, 6.5\%, and 0, respectively. $T_{\rm{c}}$ of the three crystals obtained by susceptibility measurements are below 3 K, $\sim$ 7.5 K, and $\sim$ 14.3 K, respectively (inset of Fig. 20). However, resistivities of all the three crystals show transitions at $\sim$ 14 K due to the filamentary superconductivity (Fig. 20).

Resistivities for all the three crystals maintain a nearly constant value above 150 K. From 150 K down to the superconducting transition temperature, the as-grown sample shows a nonmetallic behavior (d$\rho$/d$T <$  0), which is caused by the charge carrier localization induces by the excess Fe. This nonmetallic behavior was suppressed by removing the excess Fe and a flattened resistive behavior above $T_{\rm{c}}$ was found in the half-annealed crystal. When the excess Fe is totally removed as in the fully-annealed crystal, resistivity manifests a metallic behavior (d$\rho$/d$T >$ 0).

\begin{figure}\center	
	\includegraphics[width=8.5cm]{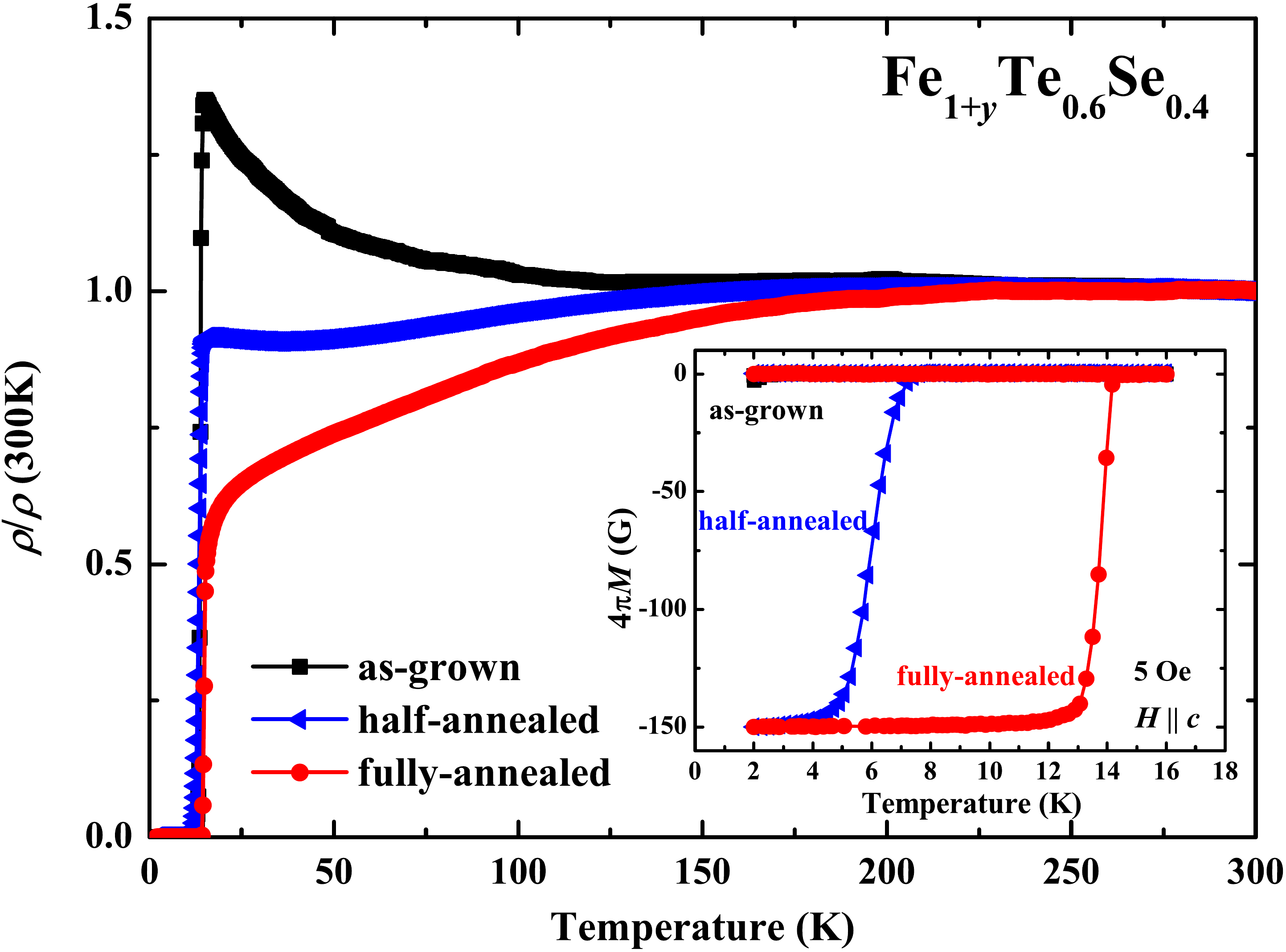}\\
	\caption{Temperature dependence of the resistivities scaled by the values at 300 K for the as-grown, half-annealed and fully-annealed Fe$_{1+y}$Te$_{0.6}$Se$_{0.4}$ single crystals. The inset shows the temperature dependences of ZFC and FZ magnetizations at 5 Oe for the three samples \cite{SunPRB}.}\label{}
\end{figure} 

Figures. 21(b)-(d) show the Hall resistivity $\rho_{yx}$ at several temperatures for the as-grown, half-annealed, and fully-annealed Fe$_{1+y}$Te$_{0.6}$Se$_{0.4}$ single crystals, respectively. The $\rho_{yx}$ for the as-grown crystal follows a linear relationship with the applied field and has a positive slope, d$\rho$$_{yx}$/d$H >$ 0. For the half-annealed crystal, $\rho_{yx}$ still keeps positive and linearly increases with magnetic field although the slopes are reduced. However, $\rho_{yx}$ of the fully-annealed crystal becomes negative when temperature decreases below 40 K, and an obvious nonlinear behavior can be witnessed. The nonlinear behavior and sign reversal observed in $\rho_{yx}$ proves the existence of multiband effect. Hall coefficients $R_H$ can be simply obtained from $R_H$ = $\rho_{yx}$/$\mu_0H$, and were shown in Fig. 21(a). For the nonlinear $\rho_{yx}$ at low temperatures in the fully-annealed crystal, $R_H$ was simply calculated from the linearly part at small fields. $R_H$ is almost temperature independent above 100 K, and keeps a constant value $\sim$1 $\times$ 10$^{-9}$ m$^3$/C for all the three samples. When temperature decreases below 100 K, an obvious divergence in $R_H$ is observed. In the as-grown crystal, $R_H$  gradually increases with decreasing temperature showing an obvious upturn at low temperatures. This upturn is almost suppressed in the half-annealed crystal, in which $R_H$ just slightly increases with decreasing temperature. In the fully-annealed crystal, $R_H$ keeps nearly temperature independence above 60 K, followed by a sudden decrease, and finally changes sign from positive to negative before approaching $T_{\rm{c}}$. The sign reversal in Hall coefficient is usually attributed to the multiband structure, indicating the dominance of electron in the charge carriers before the occurrence of superconductivity in Fe$_{1+y}$Te$_{0.6}$Se$_{0.4}$.

\begin{figure}\center	
	\includegraphics[width=8.5cm]{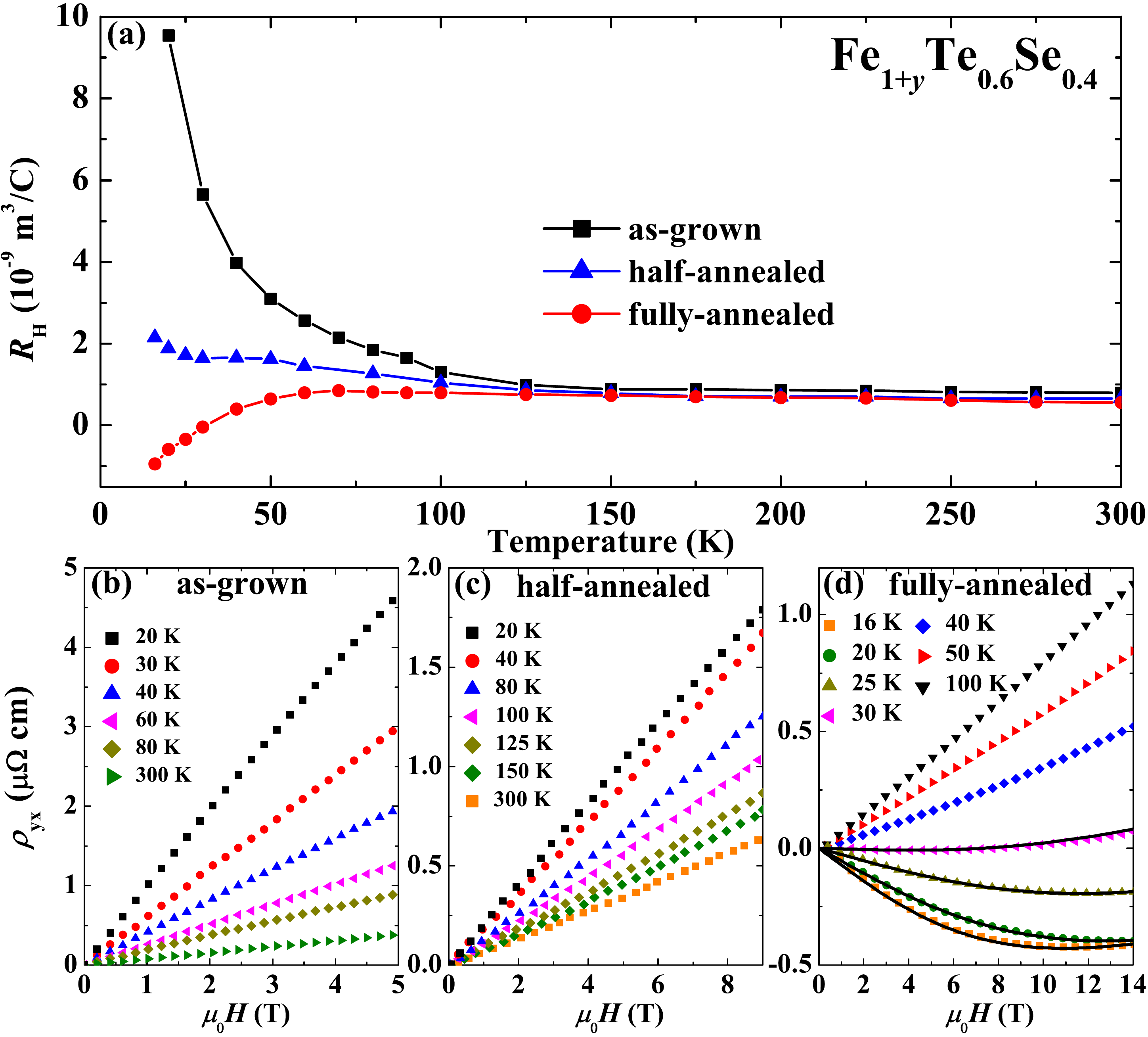}\\
	\caption{(a) Hall coefficients $R_H$ for the as-grown, half-annealed and fully-annealed Fe$_{1+y}$Te$_{0.6}$Se$_{0.4}$ single crystals. Hall resistivity $\rho_{xy}$ at several temperatures for the (b) as-grown, (c) half-annealed, and (d) fully-annealed crystals \cite{SunPRB}.}\label{}
\end{figure} 

The dramatic change in the temperature dependence of resistivity and Hall coefficients after annealing were observed in all the Fe$_{1+y}$Te$_{1-x}$Se$_x$ single crystals with different Se doping levels as shown in Fig. 22 \cite{SunPDSciRep}.    

\begin{figure}\center	
	\includegraphics[width=14cm]{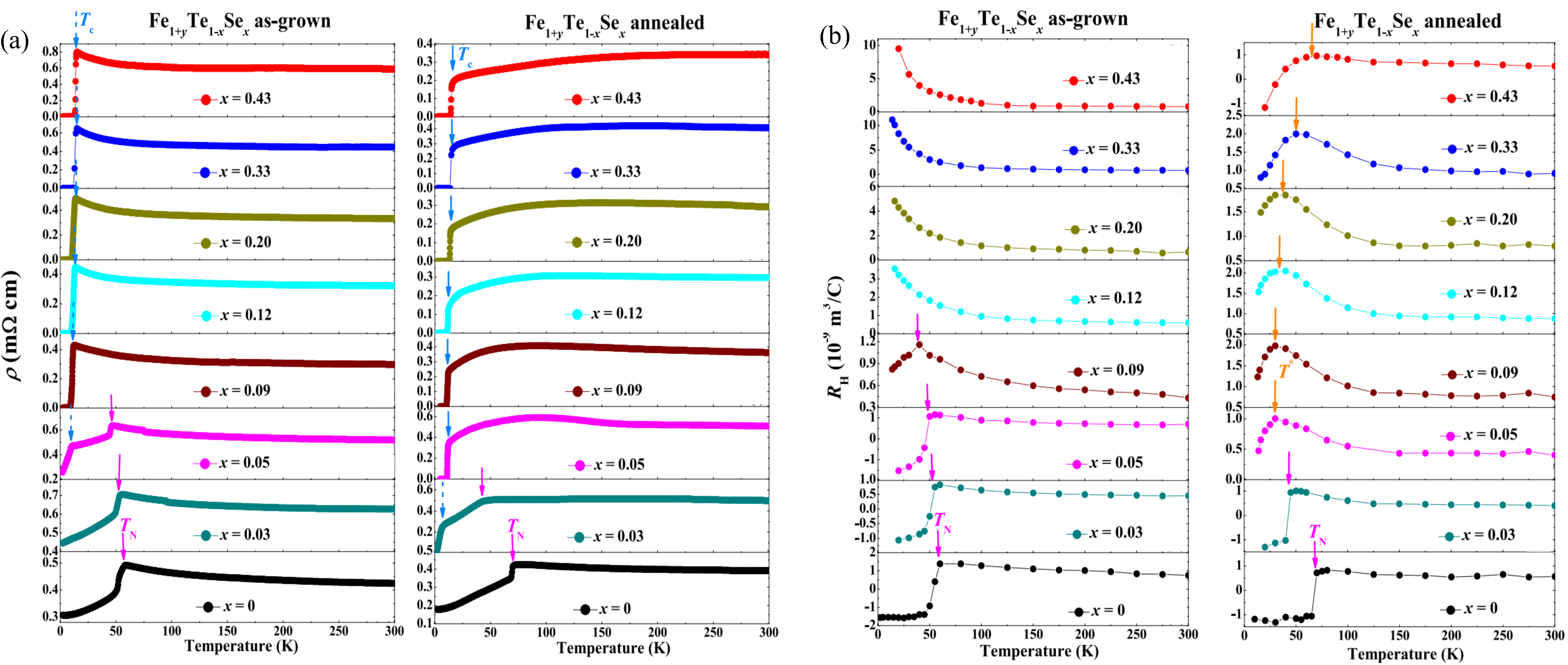}\\
	\caption{Temperature dependence of (a) in-plane resistivity, and (b) Hall coefficients for the as-grown (left panel) and O$_2$-annealed (right panel) Fe$_{1+y}$Te$_{1-x}$Se$_x$ (0 $\leq$ $x$ $\leq$ 0.43) single crystals \cite{SunPDSciRep}.}\label{}
\end{figure} 

In a multiband system, the MR is usually described by the following expression \cite{EomPRB},
\begin{equation}
\label{eq.2}
MR\equiv\frac{\Delta\rho(H)}{\rho(0)}\approx\frac{1}{2}\frac{\Sigma_i\Sigma_{j\neq i}\sigma_i\sigma_j(\omega_{ci}\tau_i-\omega_{cj}\tau_j)^2}{(\Sigma_i\sigma_i)^2},
\end{equation}
where $\sigma_i$ is the conductivity, $\tau_i$ is the relaxation time and $\omega_{ci}$ is the cyclotron frequency, which has opposite sign for electron and hole bands. In this case, the ($\omega_{ci}$$\tau_i$-$\omega_{cj}$$\tau_j$)$^2$ term becomes larger because the $\omega_{ci}$$\tau_i$ terms add up, which will results in a large positive MR. However, the MR of the as-grown crystal is just $\sim$ 0.03\% at 16 K under 9 T as shown in the inset of Fig. 23(a). This unexpected small MR can be also explained by the effect of excess Fe. Recent neutron scattering measurements revealed that the excess Fe induces a magnetic Friedel-like oscillation at ($\pi$, 0) order and involves more than 50 neighboring Fe sites \cite{ThampyPRL}. Spins from those Fe clusters will be weakly polarized under magnetic field inducing a negative MR, which will cancel out the positive MR of the sample itself. Actually, previous reports on Fe$_{1+y}$Te$_{1-x}$Se$_x$ all show such small values of MR, and sometimes even negative MR was observed \cite{ChangSUST,HuPRB}. Such small MR is increased to $\sim$ 0.14\% in the half-annealed crystal because parts of the excess Fe was removed. After totally removing the excess Fe, MR of the fully-annealed Fe$_{1+y}$Te$_{0.6}$Se$_{0.4}$ reaches larger than 17\% at 16 K under 14 T. 

More interesting, MR of the fully-annealed Fe$_{1+y}$Te$_{0.6}$Se$_{0.4}$ linearly increases with the applied field from intermediate field (e. g. 2 T at 16 K) to the measurement limit of 14 T, whereas a small parabolic bend just remains at low fields. This is in sharp contrast to the semi-classical field dependence of MR, in which MR generally develops in proportion to $H^2$ over the entire field range. The linear dependence of MR on field is more evident in the first-order derivative d$MR$/d$B$ as shown in Fig. 23(b). d$MR$/d$B$ is proportional to magnetic field at low $H$, then saturated at high fields. The linear MR can be interpreted by considering a quantum limit where all the carriers occupy only the lowest Laudau level (LL)\cite{AbrikosovPRB,AbrikosovEPL}. This situation usually happens when the field is very large and the difference between the zeroth and first Landau levels $\Delta_{LL}$ exceeds the Fermi energy $E_F$ and the thermal fluctuation $k_BT$. However, the linear MR was identified in low field region in some materials hosting Dirac fermions with linear energy dispersion, such as graphene \cite{NovoselovNature}, topological insulators \cite{TaskinPRL}, Ag$_{2-\delta}$(Te/Se) \cite{XuNature}, $\alpha$-(BEDT-TTF)$_2$I$_3$ \cite{Kobayashijpsj}, some layered compounds with two-dimensional Fermi surface (like SrMnBi$_2$ \cite{ParkPRL,WangPRB}) and iron-based Ba(Sr)Fe$_2$As$_2$ \cite{RichardPRL,HuynhPRL,ChongEPL}  and La(Pr)FeAsO \cite{PallecchiPRB,BhoiPrFeAsO}. For the Dirac state, $\Delta_{LL}$ is described as $\Delta_{LL}$=$\pm$$v_F$$\sqrt{2e\hbar B}$, leading to a much larger LL splitting compared with the parabolic band. Consequently, the quantum limit  can be achieved in low field region \cite{AbrikosovPRB}.

Characteristic field $B^*$, defined as the crossover field between the semiclassical regime and the quantum linear regime, is marked by the arrow in Fig. 23(b). The temperature dependence of $B^*$ is shown in Fig. 23(c), which is obviously violating the linear relation expected from conventional parabolic bands, and can be well fitted by $B^*$ = (1/2e$\hbar$$v^2_F$)(k$_BT$+$E_F$)$^2$ for the Dirac fermions as shown in Fig.23 (c) \cite{TaskinPRL}. The good agreement of $B^*$ with the above equation indicates the existence of Dirac fermions in Fe$_{1+y}$Te$_{0.6}$Se$_{0.4}$. Recently, the Dirac cone band structure and the topological superconductivity in FeTe$_{1-x}$Se$_x$ were confirmed by the ARPES measurements \cite{ZhangARPESScience,ZhangPengNatPhy}. The report in \cite{ZhangPengNatPhy} was performed on our fully-annealed single crystals. The fitting gives a large Fermi velocity $v_F$ $\sim$ 1.1 $\times$ 10$^5$ ms$^{-1}$ and $E_F$ $\sim$ 5.5 meV, which are close to the previous reports in similar compounds BaFe$_2$As$_2$ ($v_F$ $\sim$ 1.88 $\times$ 10$^5$ ms$^{-1}$, $E_F$ $\sim$ 2.48 meV) \cite{HuynhPRL} and SrFe$_2$As$_2$ ($v_F$ $\sim$ 3.11 $\times$ 10$^5$ ms$^{-1}$ and $E_F$ $\sim$ 6.9 meV) \cite{ChongEPL}. 

\begin{figure}\center	
	\includegraphics[width=8.5cm]{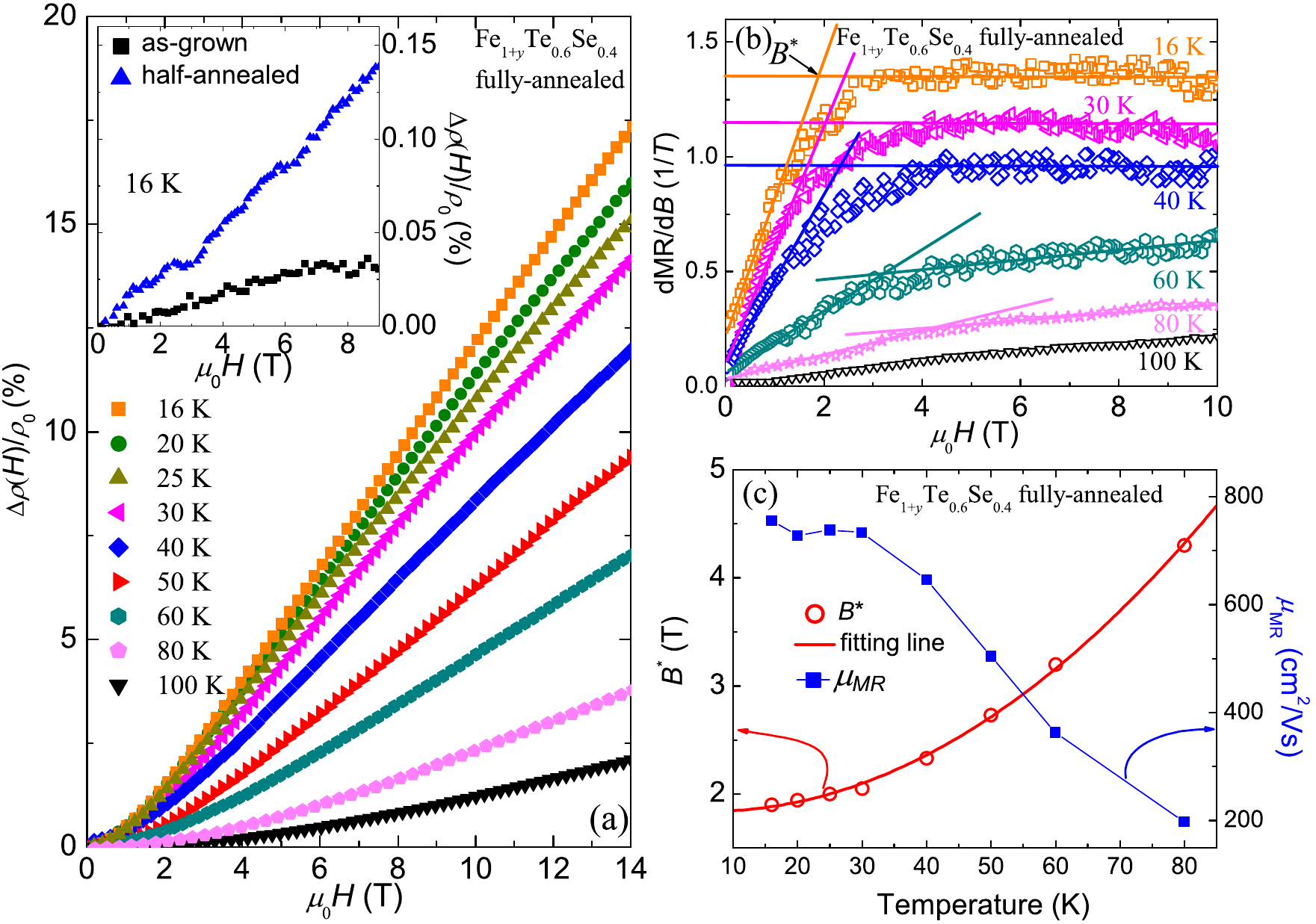}\\
	\caption{(a) Magnetic field dependence of  magnetoresistance (MR=($(\rho(H)-\rho(0))/\rho(0)$)) for the fully-annealed Fe$_{1+y}$Te$_{0.6}$Se$_{0.4}$ single crystal at different temperatures. Inset is the MR for the as-grown and half-annealed crystals at 16 K. (b) The field derivative of in-plane MR at different temperatures for the fully-annealed Fe$_{1+y}$Te$_{0.6}$Se$_{0.4}$ single crystal. The solid lines denote the semiclassical regime and the quantum linear region, respectively. The characteristic field $B^*$ is marked by the arrow. (c) Temperature dependence of the characteristic field $B^*$ (red circles) and the effective MR mobility $\mu_{MR}$ (blue squares). Red line is the fitting of $B^*$ by $B^*$ = (1/2e$\hbar v^2_F$)(k$_BT$+$E_F$)$^2$ \cite{SunPRB}.}\label{}
\end{figure}

\subsection{Annealing effects on band structure}

The large difference in Hall effects of the as-grown, half-annealed, and fully-annealed crystals indicates that the band structure should be also affected by the excess Fe, i.e. changing after annealing. Details about the band structure evolution with decreasing of excess Fe were observed by the ARPES measurement (see Figs. 24 (A-F) \cite{ShaharARPESSciAdv}, and the schematics of the electronic structures in Fig. 24(G). With decreasing excess Fe, the electron-type band at $M$-point, and the hole-type $\alpha_2$-band at the $\Gamma$ point move down, which is consistent with the Hall effect in Fig. 21(a). In contrast to the electron and $\alpha_2$-band, no significant shift in hole $\alpha_1$ and $\alpha_3$ band was observed. More interesting, the Fermi energy $\epsilon_{\rm{F}}$ obtained from the $\alpha_2$ hole band decreases from 19 meV to 6 meV with decreasing excess Fe, while the SC gap $\Delta$ keeps $\sim$ 3 meV. Thus, $\Delta$/$\epsilon_{\rm{F}}$ increases from $\sim$ 0.16 to $\sim$ 0.5 after decreasing excess Fe, which indicates that the Bardeen-Cooper-Schrieffer (BCS) to BoseEinstein-condensate (BEC) crossover \cite{Lubashevskynatphy} can be tubed by excess Fe.        

\begin{figure}\center	
	\includegraphics[width=14cm]{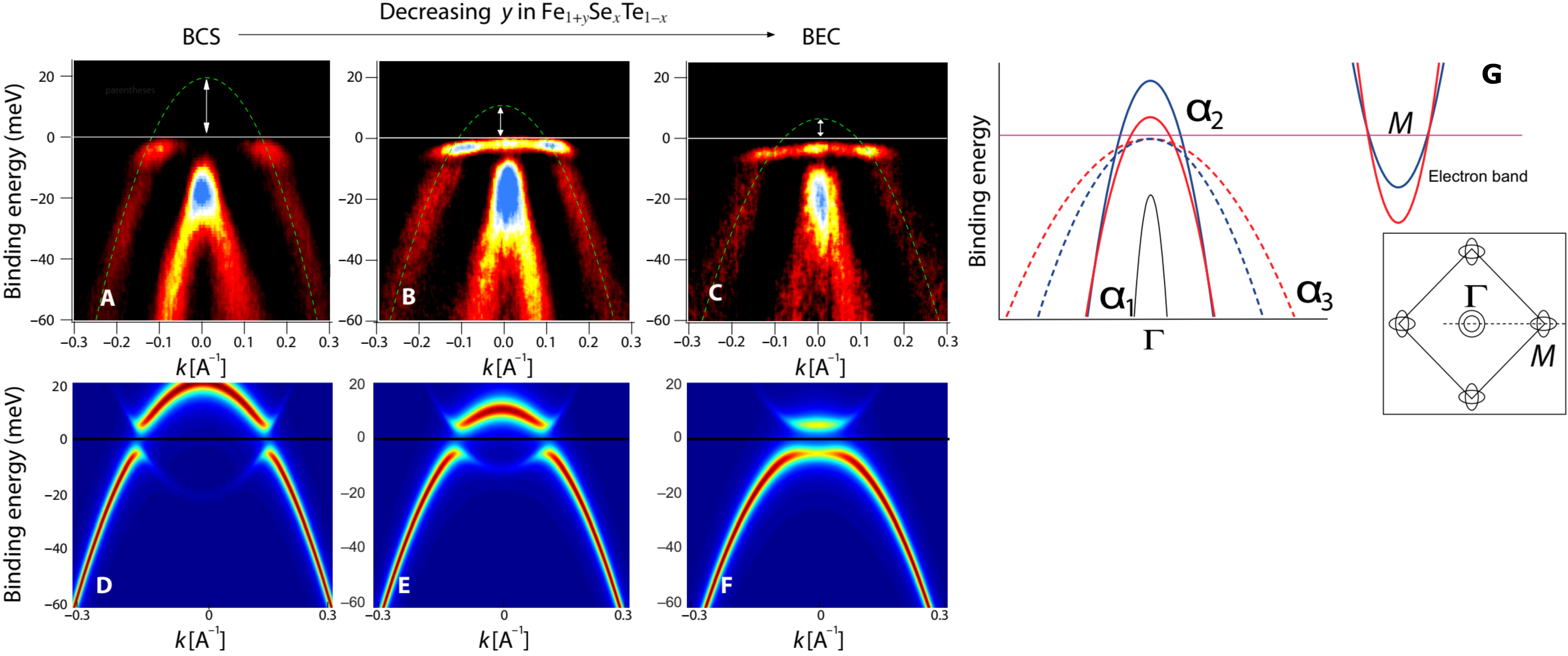}\\
	\caption{(A to C) ARPES spectra for three Fe$_{1+y}$Te$_{1-x}$Se$_x$ samples in order of decreasing $y$ (excess Fe) from left to right. The spectra are normalized using the intensity from high-order photons, and a constant background is removed. The spectra are sharpened by adding a small part of their second derivative to the original data. The green dashed line is the best fit to the data using a simple parabolic dispersion. (D to F) Spectral functions, calculated using the model and parameters described in the text, to describe the BCS-BEC crossover seen in the data in the top panels.(G) Schematic band structure and the effect of excess Fe on various bands. The red and blue curves correspond to the samples with small and large amount of excess Fe, respectively. Reprinted with permission from \cite{ShaharARPESSciAdv}. Copyright 2017 by the American Association for the Advancement of Science: Science Advances.}\label{}
\end{figure} 

\section{Annealing effects on superconducting properties} 

\subsection{Annealing effects on $T_{\rm{c}}$ and phase diagram}

The annealing effect on $T_{\rm{c}}$ has already been systematically studied (see Fig. 3, Fig. 4, and Fig. 6). To check the bulk nature of the SC, we also performed specific heat measurements on both the as-grown and fully-annealed Fe$_{1+y}$Te$_{0.6}$Se$_{0.4}$ single crystal (Fig. 25) \cite{SunSciRep}.  Obviously, no specific heat jump can be observed in the as-grown crystal, which proves that the superconducting transition observed in the resistivity measurement (see Fig. 20) comes from the filamentary superconductivity. After annealing, a clear specific heat jump can be observed, indicating the bulk nature of the superconductivity. Specific heat jump $\Delta C/T_{\rm{c}}$ was obtained as $\sim$ 66.3 mJ/molK$^2$ with $T_{\rm{c}}$ = 14.5 K determined by entropy balance consideration. The normalized specific heat jump at $T_{\rm{c}}$, $\Delta C/\gamma_{\rm{n}}T_{\rm{c}}$, is $\sim$ 3.0 by using the fitted Sommerfeld coefficient of $\gamma_n$ = 22 mJ/molK$^2$. The value is much larger than BCS weak-coupling value of 1.43 indicating that superconductivity in the fully-annealed Fe$_{1+y}$Te$_{0.6}$Se$_{0.4}$ is strong coupling in nature. The specific heat evidences for the bulk SC were also reported in the annealed Fe$_{1+y}$Te$_{1-x}$Se$_x$ with $x >$ 0.05 \cite{NojiJPSJPhaseD}.

\begin{figure}\center	
	\includegraphics[width=8.5cm]{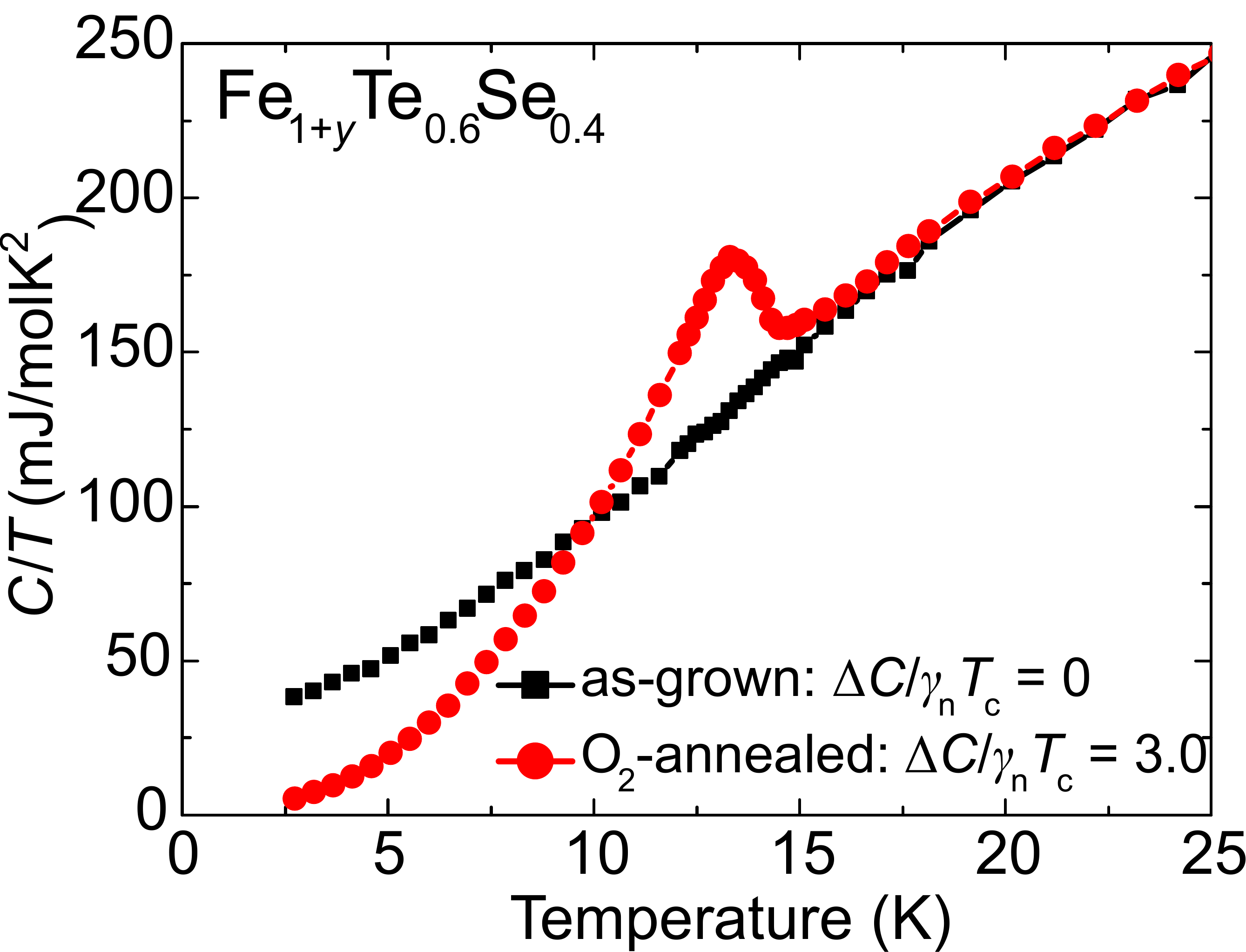}\\
	\caption{Temperature dependence of specific heat plotted as $C/T$ vs $T$ for the as-grown and O$_2$-annealed Fe$_{1+y}$Te$_{0.6}$Se$_{0.4}$ single crystals \cite{SunSciRep}.}\label{}
\end{figure} 
 
Phase diagrams of Fe$_{1+y}$Te$_{1-x}$Se$_x$ have been already reported by several groups \cite{LiuNatMat,NojiFeTeSeannealing,DongPRB,KatayamaPhaseD,KawasakiSSC}. However, they are all under debate. Some basic information is even controversial in those reported results, like the region of bulk SC, the coexistence of AFM and SC, and the spin glass state. Two well-known phase diagrams are shown in Figs. 26(a) and (b) \cite{LiuNatMat,KatayamaPhaseD}. In Fig. 26(a), bulk SC was observed only in the region of $x >$ 0.27, and the coexistence of AFM and SC was reported. In the phase diagram of Fig. 26(b), spin glass state was reported together with the superconductivity. Based on the above discussion, these controversies come from the sample-dependent amount of excess Fe. The phase diagram of Fe$_{1+y}$Te$_{1-x}$Se$_x$ was also revised after reducing excess Fe by annealing in vacuum \cite{NojiJPSJPhaseD}, air \cite{DongPRB}, and O$_2$ \cite{KawasakiSSC}. However, due to the remaining of excess Fe as well as the lack of bulk evidences for the SC, the phase diagrams are still not unified.     

Based on the description in the experimental part, the excess Fe was confirmed to be totally removed in our crystal after careful annealing. Thus, we established a doping-temperature ($x-T$) phase diagrams ((Figs. 26(c) and (d)) for both the as-grown and fully-annealed (without excess Fe, $y$ = 0) Fe$_{1+y}$Te$_{1-x}$Se$_x$ based on the magnetic susceptibility (Fig. 19), resistivity (Fig. 22(a)), and Hall effect measurements (Fig. 22(b)) shown above \cite{SunPDSciRep}. For the as-grown crystals, in the doping region of $x <$ 0.12, the AFM transition, $\sim$ 58 K in the non-doped Fe$_{1+y}$Te, is monotonically suppressed with increasing Se substitution. Accompanied by the suppression of AFM, SC emerges from $x$ = 0.05, and coexists with the antiferromagnetic phase until $x <$ 0.13. That SC, marked by the squares, is not bulk in nature, and can be only observed in resistivity measurement. For $x \geq$ 0.12, the AFM transition is absent and replaced by a spin glass state (observed by magnetic susceptibility measurements, and marked by the right-triangles) similar to Fig. 26(b). The spin glass state may be originated from the effect of excess Fe, which interacts with more than 50 neighboring Fe in the adjacent Fe layers, and induces the magnetic Friedel-like oscillation \cite{ThampyPRL}.
 
After removing the excess Fe by annealing, the phase diagram of Fe$_{1+y}$Te$_{1-x}$Se$_x$ is dramatically changed. As shown in Fig. 26(d), the AFM state is suppressed into a very narrow region of $x <$ 0.05, and the spin glass state is disappeared. Immediately after the disappearance of AFM state, bulk SC emerges, and is observed in the doping region of $x \geq$ 0.05. The coexistence of AFM and SC states is absent in the annealed crystals. Thus, the previously reported coexistence of AFM, spin glass state with SC is originated from the effect of excess Fe. Besides, the characteristic temperature $T^*$ observed in the $R_H$ (orange arrows in Fig. 22(b)) is plotted in the figure, which also resides in the doping region of $x \geq$ 0.05, and gradually increases with increasing Se doping. It suggests that the multiband effect in Fe$_{1+y}$Te$_{1-x}$Se$_x$ may be strongly related to the occurrence of SC. On the other hand, the rapid suppression of AFM state with a small amount of doping, absence of coexistence of the AFM and SC states are all similar to the phase diagrams of LaFeAsO$_{1-x}$F$_x$ \cite{LuetkensPDNatMat} and CeFeAsO$_{1-x}$F$_x$ \cite{ZhaoPDNAtMat}. The step-like behavior of the magnetism and SC in the small region of 0.03 $< x <$ 0.05 suggests that the SC in the Fe$_{1+y}$Te$_{1-x}$Se$_x$ system may be related to the suppression of static magnetic order rather than the increase of the effective charge carrier density by the doping or structural distortion.

\begin{figure}\center	
	\includegraphics[width=14cm]{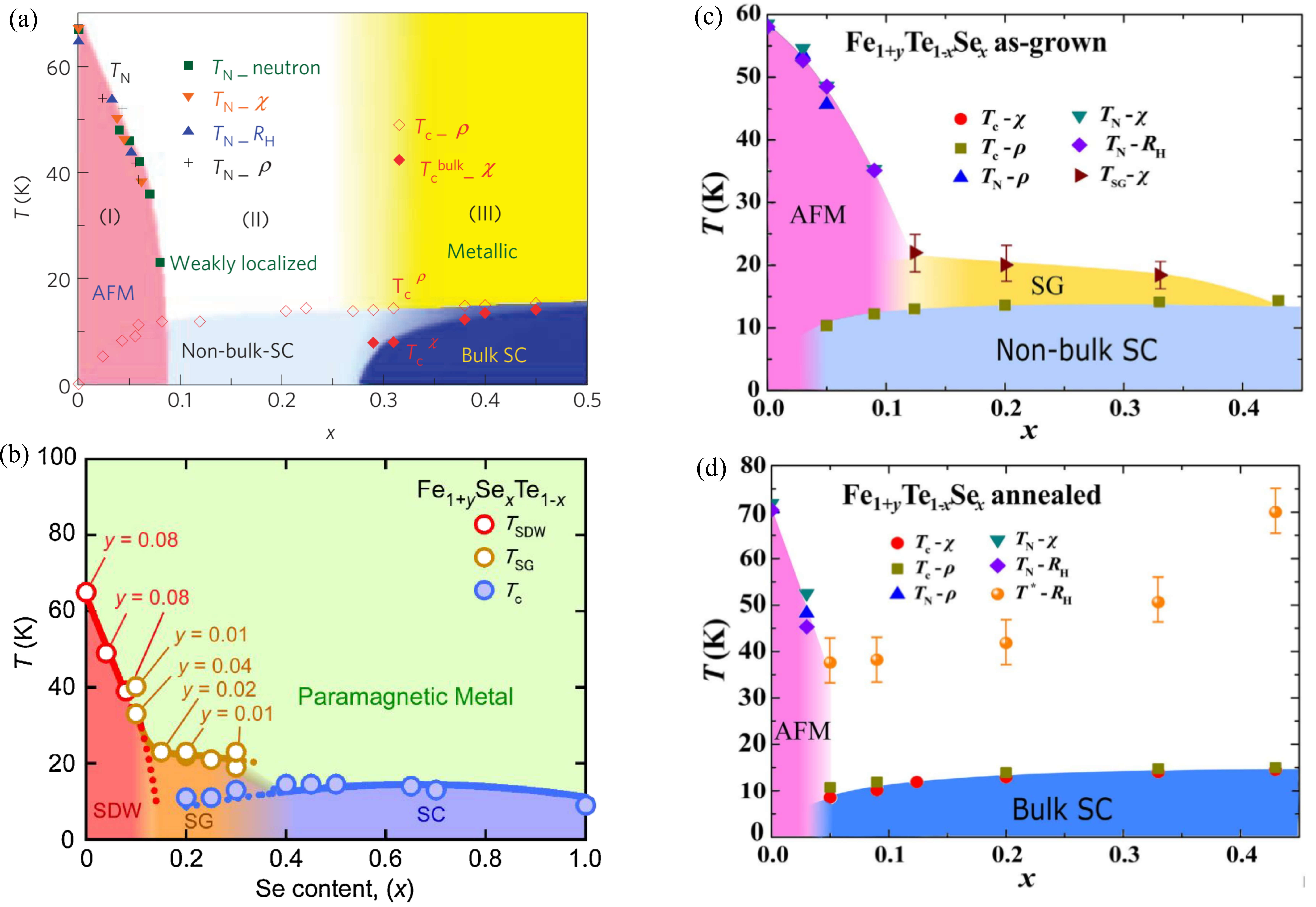}\\
	\caption{The doping-temperature ($x-T$) phase diagrams for Fe$_{1+y}$Te$_{1-x}$Se$_x$ single crystals reported by (a) \cite{LiuNatMat}, (b) \cite{KatayamaPhaseD}, and the phase diagrams for our (c) as-grown and (d) annealed single crystals \cite{SunPDSciRep}. (a) Reprinted with permission from \cite{LiuNatMat}. Copyright 2010 by the MacMillan Publishers Ltd: Nature Materials. (b) Reprinted with permission from \cite{KatayamaPhaseD}. Copyright 2010 by the Physical Society of Japan.}\label{}
\end{figure} 

\subsection{Annealing effects on upper critical field and anisotropy}

To study the effect of annealing on the upper critical filed, the resistive transitions measured in magnetic field up to 9 T for $H \parallel c$ and $H \parallel ab$ for both the as-grown and annealed Fe$_{1+y}$Te$_{0.6}$Se$_{0.4}$ single crystals were compared in Fig. 27. For both crystals, with increasing field along $c$-direction, the resistive transition shifts to lower temperatures accompanied by a slight increase in the transition width, while this broadening is almost negligible for $H \parallel ab$. Compared with the as-grown crystal, the resistive transition shift under magnetic field becomes weaker in the annealed one. The upper critical fields $H_{c2}$ for $H \parallel c$ and $H \parallel ab$ defined by the 90\% of resistive transition are plotted in Fig. 27(e). The anisotropy $\gamma$ defined as the ratio of $H_{c2}^{ab}$/$H_{c2}^c$ close to $T_{\rm{c}}$ was shown in the inset of Fig. 27(e). Obviously, $\gamma$ for the as-grown crystal is $\sim$ 2.5 - 3, which is slightly larger than the value $\sim$ 1.5 - 2 for the annealed crystal.      

\begin{figure}\center	
	\includegraphics[width=8.5cm]{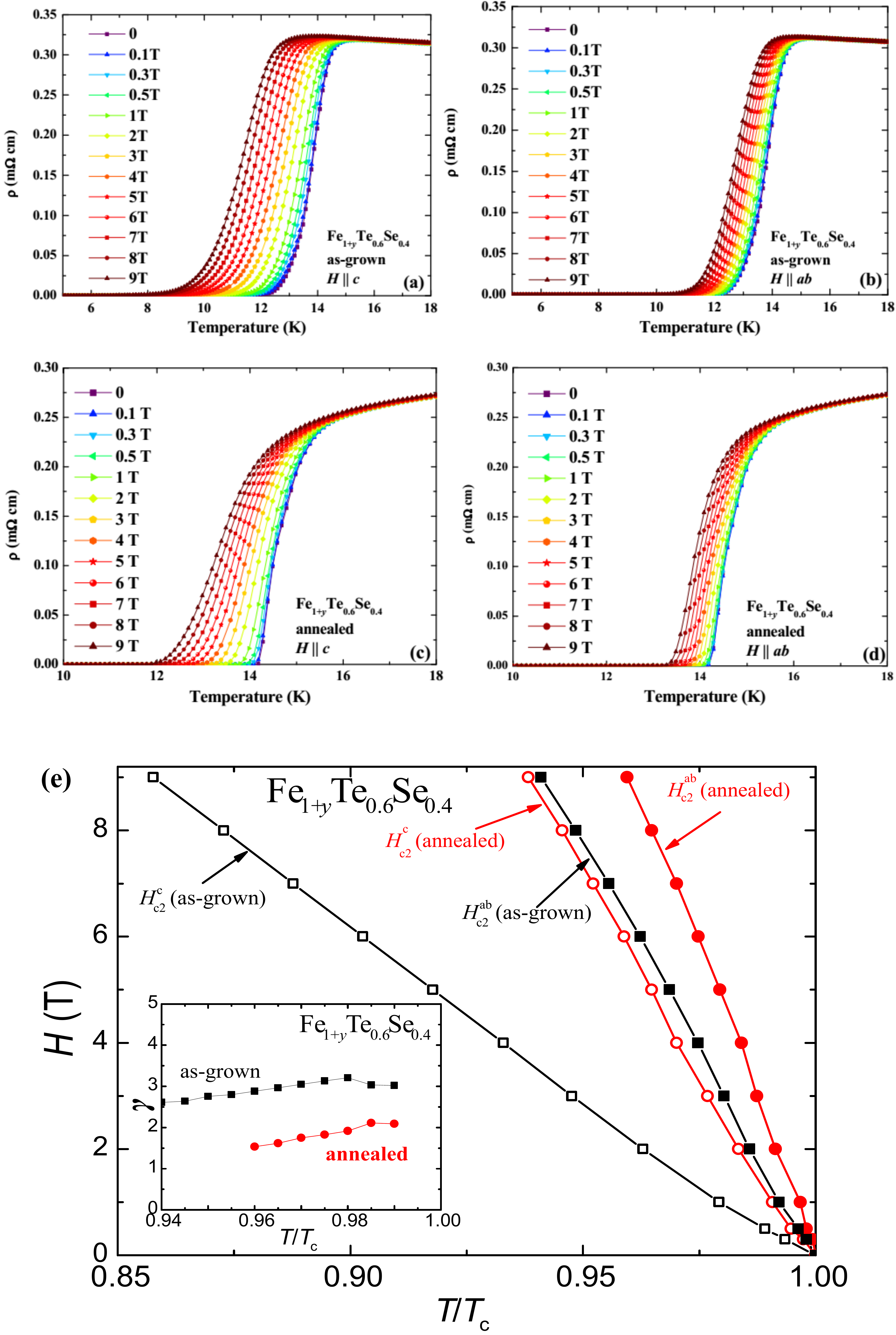}\\
	\caption{The resistive transitions of (a-b) as-grown and (c-d) fully-annealed Fe$_{1+y}$Te$_{0.6}$Se$_{0.4}$ single crystals measured in magnetic field up to 9 T for $H \parallel c$ and $H \parallel ab$. (e) Temperature dependence of upper critical fields for $H \parallel c$ and $H \parallel ab$. The $H_{c2}$ is defined by the 90\% of resistive transition. Inset shows the anisotropy $\gamma$ close to $T_{\rm{c}}$ for the as-grown and fully-annealed crystals.}\label{}
\end{figure} 

To get more information about the $H_{c2}$ and $\gamma$ at low temperatures, we also measured the resistive transitions under pulse magnetic field up to 50 T \cite{Paninprepare}. The $H_{c2}$ for the as-grown, half-annealed, and fully-annealed crystals were shown in Figs. 28(a)-(c), respectively. All the three crystals show large value of $H_{c2}$ $\sim$ 45 - 50 T at 2 K, which is not affected by the excess Fe. On the other hand, $\gamma$ is decreased with reducing the amount of excess Fe in a large temperature range (see Fig. 28(d)), which is similar to the results close to $T_{\rm{c}}$ shown in the inset of Fig. 27(e). At low temperature of $T/T_{\rm{c}} \leq$ 0.3, all the three crystals manifest a small $\gamma$ $\sim$ 1.          

\begin{figure}\center	
	\includegraphics[width=8.5cm]{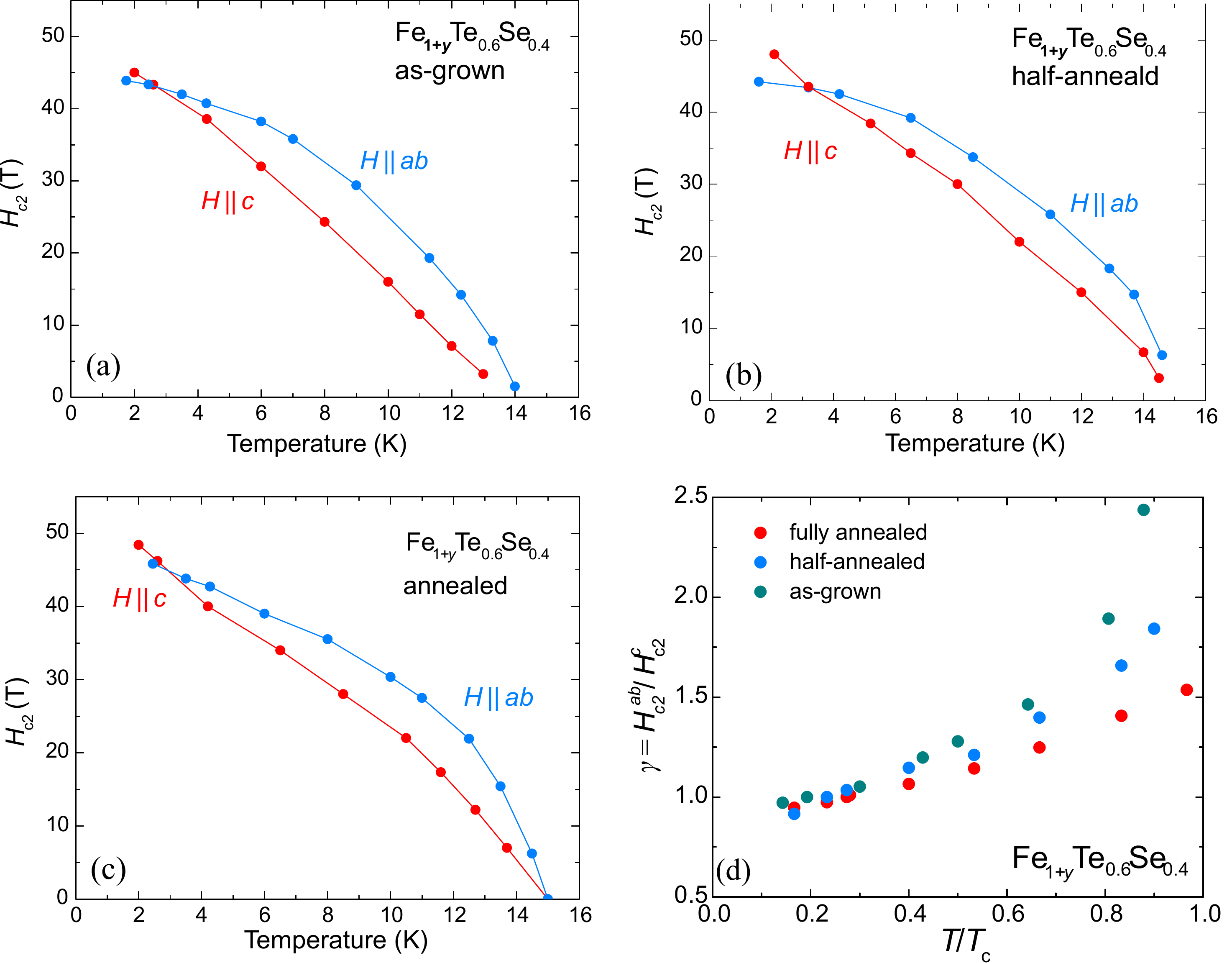}\\
	\caption{(a-c) Temperature dependence of upper critical fields for $H \parallel c$ and $H \parallel ab$ of the as-grown, half-annealed, and the fully-annealed Fe$_{1+y}$Te$_{0.6}$Se$_{0.4}$ single crystals. (d) Anisotropy $\gamma$ at the temperature rang of 0.1 $\leq$ $T/T_{\rm{c}}$ $\leq$ 1 for the three crystals \cite{Paninprepare}.}\label{}
\end{figure} 

The excess Fe was also reported to affect the anisotropy in the $ab$-plane \cite{LiuinplaneannisotropyPhysRevB.91.134502}. Fig. 29(a) shows that $\rho_a$ and $\rho_b$ are almost identical at temperature above $T_{\rm{N}}$, while become different in the AFM state. The difference between $\rho_a$ and $\rho_b$ is decreased with reducing the excess Fe. Similar behavior was also observed in the Se-doped FeTe (see Fig. 29(b)). However, in Fe$_{1+y}$Te$_{1-x}$Se$_x$, $\rho_a > \rho_b$, which is opposite to that of $\rho_a < \rho_b$ observed in IBSs-122 samples \cite{ChuScience824}.         

\begin{figure}\center	
	\includegraphics[width=8.5cm]{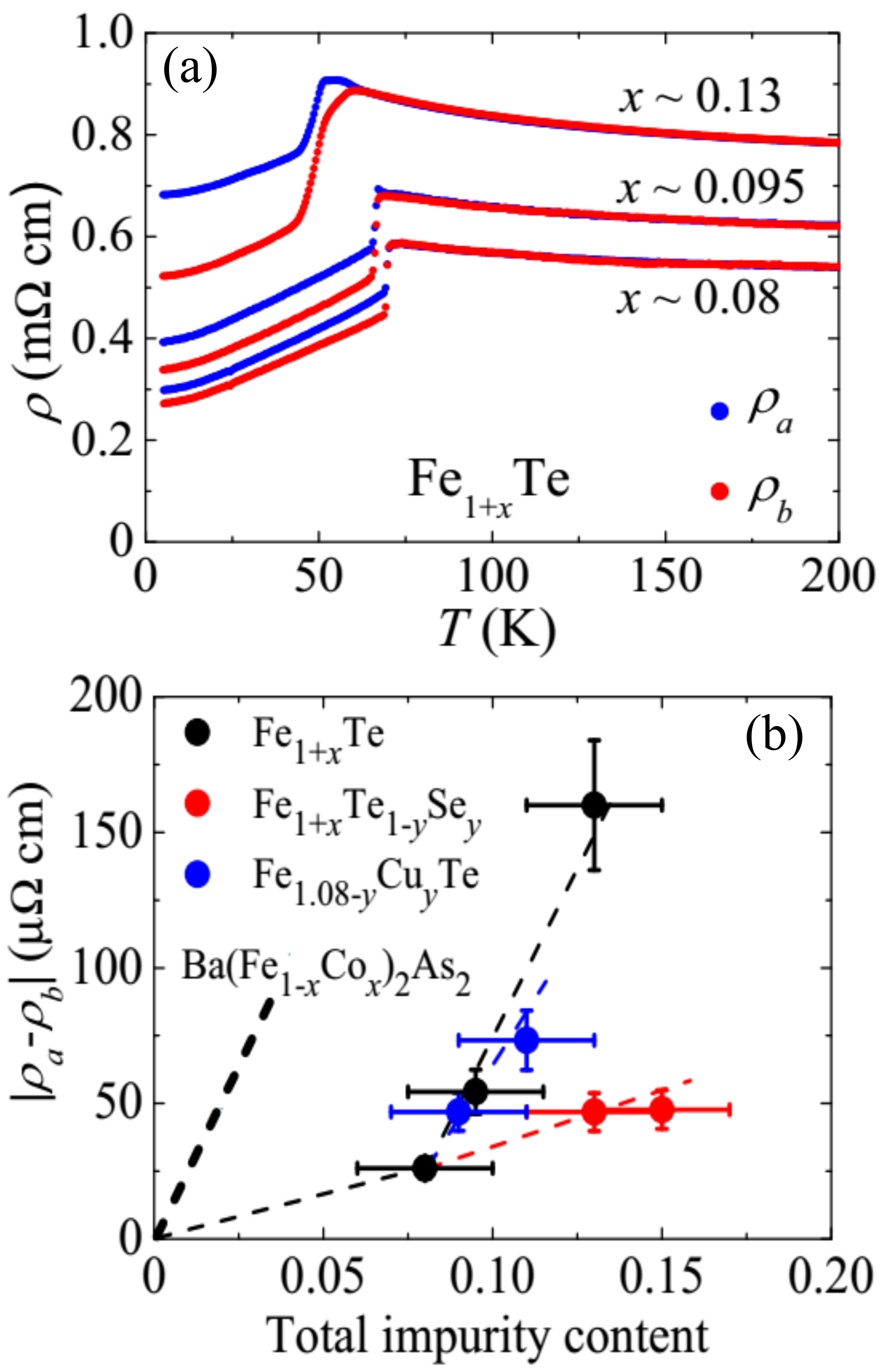}\\
	\caption{(a) Temperature dependence of the in-plane	resistivity anisotropy measured on detwinned Fe$_{1+x}$Te crystals with	three different excess Fe contents. (b) In-plane resistivity anisotropy in the residual component ($\mid\rho_a - \rho_b\mid$) plotted against the total impurity content:	$x$ for Fe$_{1+x}$Te, ($x$ + $y$) for Fe$_{1+x}$Te$_{1-y}$Se$_y$ and (0.08 + $y$) for Fe$_{1.08-y}$Cu$_y$Te \cite{LiuinplaneannisotropyPhysRevB.91.134502}. The thick dashed line illustrates the dependence of the resistivity anisotropy on the substituted Co content $x$ for Ba(Fe$_{1-x}$Co$_x$)$_2$As$_2$, reproduced from Ref. \cite{IshidaNematic122PhysRevLett.110.207001}. Reprinted with permission from \cite{LiuinplaneannisotropyPhysRevB.91.134502}. Copyright 2015 by the American Physical Society.}\label{}
\end{figure} 

\subsection{Annealing effects on critical current density}

As discussed above, the excess Fe affects the superconductivity of Fe$_{1+y}$Te$_{1-x}$Se$_x$. Hence, it will also affect the $J_{\rm{c}}$ of the crystal. Earlier studies on crystals with different amount of excess Fe reported a quite different values of $J_{\rm{c}}$ at the same situation (such as 5 K, zero field), ranging from 10$^4$ A/cm$^2$ to over 10$^5$ A/cm$^2$ \cite{TaenPRB,BonuraJcPhysRevB.85.134532,DasJcPhysRevB.84.214526,YadavJcNJP}. To systematically study the annealing effect on the $J_{\rm{c}}$, we also measured the magnetic hysteresis loops (MHLs) after the susceptibility in the experiment of annealing  Fe$_{1+y}$Te$_{0.6}$Se$_{0.4}$ single crystals with increasing amount of O$_2$ (susceptibility results are shown in Fig. 3). $J_{\rm{c}}$ can be calculated by using the extended Bean model \cite{Beanmodel}:
\begin{equation}
\label{eq.1}
J_{\rm{c}}=20\frac{\Delta M}{a(1-a/3b)},
\end{equation}
where $\Delta$\emph{M} is \emph{M}$_{\rm{down}}$ - \emph{M}$_{\rm{up}}$, \emph{M}$_{\rm{up}}$ [emu/cm$^3$] and \emph{M}$_{\rm{down}}$ [emu/cm$^3$] are the magnetization when sweeping fields up and down, respectively, \emph{a} [cm] and \emph{b} [cm] are sample widths (\emph{a} $<$ \emph{b}). Self-field $J_{\rm{c}}$ at 2 K vs. the amount of O$_2$ are summarized in Fig. 30 together with the change in $T_{\rm{c}}$. $J_{\rm{c}}$ are gradually enhanced together with the increase of $T_{\rm{c}}$ by annealing with increasing amount of O$_2$, and reach the values larger than 10$^5$ A/cm$^2$. 

\begin{figure}\center	
	\includegraphics[width=8.5cm]{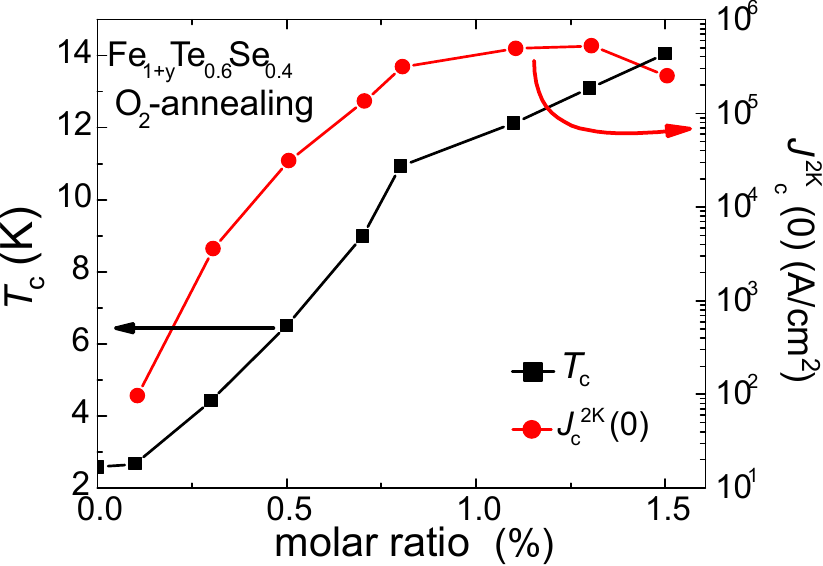}\\
	\caption{Changes in $T_{\rm{c}}$ and self-field $J_{\rm{c}}$ (2 K) for Fe$_{1+y}$Te$_{0.6}$Se$_{0.4}$ single crystal annealed at 400$^\circ$C with increasing amount of O$_2$ \cite{SunSciRep}.}\label{}
\end{figure} 

Figure 31 shows the magnetic field dependences of $J_{\rm{c}}$ for $H \parallel c$ and $H \parallel ab$ at different temperatures for the O$_2$-annealed Fe$_{1+y}$Te$_{1-x}$Se$_x$ ($x$ = 0.1, 0.2, 0.3, and 0.4) \cite{SunAPRE,SunEPL}. For all the crystals, $J_{\rm{c}}$ reaches a large value over 10$^5$ A/cm$^2$ at 2 K (self-field) for both $H \parallel c$ and $H \parallel ab$. Crystals annealed in other atmospheres such as the Te, Se, S, P, Sb, As, and I all show similar values of $J_{\rm{c}}$ \cite{Sunjpsj,SunJPSJshort,YamadaJPSJPanneal,ChenI2annealJPSJ,ZhouweiAsanneal,TamegaiJcannealed}. The $J_{\rm{c}}$ of Fe$_{1+y}$Te$_{1-x}$Se$_x$ is the highest among iron chalcogenides \cite{SunFeJcPhysRevB.92.144509,HuFeTeSJcPhysRevB.80.214514,LeiHechangPhysRevB.83.184504,SunLiFeOHFeSeJc}.     

\begin{figure}\center	
	\includegraphics[width=16cm]{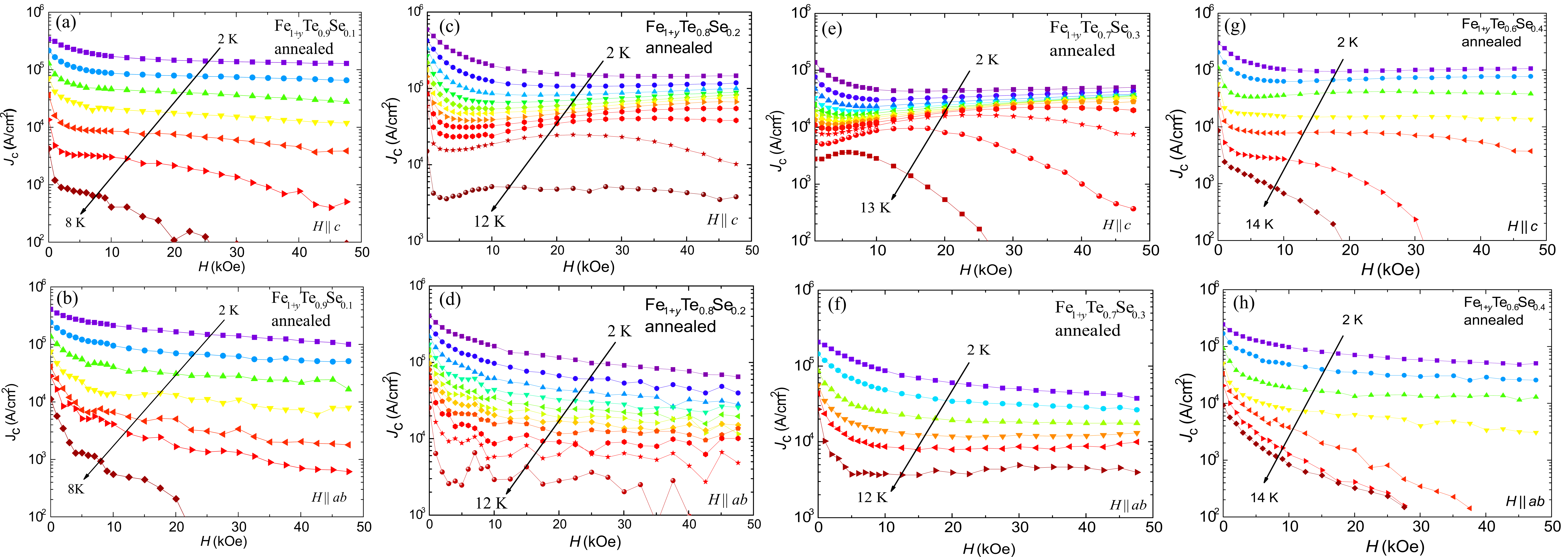}\\
	\caption{Magnetic field dependences of critical current densities for $H \parallel c$ (up panel) and $H \parallel ab$ (lower panel) at different temperatures for the O$_2$-annealed (a-b) Fe$_{1+y}$Te$_{0.9}$Se$_{0.1}$, (c-d) Fe$_{1+y}$Te$_{0.8}$Se$_{0.2}$, (e-f) Fe$_{1+y}$Te$_{0.7}$Se$_{0.3}$, and (g-h) Fe$_{1+y}$Te$_{0.6}$Se$_{0.4}$ \cite{SunSUSTevolution}.}\label{}
\end{figure} 

The above estimations of $J_{\rm{c}}$ using the Bean model rely on the assumption that homogeneous currents flow within the sample. To examine this assumption, we took MO images on the annealed Fe$_{1+y}$Te$_{1-x}$Se$_x$ ($x$ = 0.1, 0.2, 0.3, and 0.4) single crystals in the remanent state, which was prepared by applying 800 Oe along the $c$-axis for 1 s and removing it after zero-field cooling. Typical MO images for the four crystals are shown in Figs. 32(a)-(d) \cite{SunSUSTevolution}. All the MO images manifest typical roof-top patterns, indicating a nearly uniform current flow in the $ab$-plane of the crystals. Besides, the typical current discontinuity lines (so-called d-line), which cannot be crossed by vortices, can be observed in all the crystlas and marked by the dashed lines in Figs. 32(a)-(d). By measuring the angles of the discontinuity line for our rectangular sample, the in-plane anisotropy of the current densities can be easily estimated. For all the crystals, the angle $\theta$ is $\sim$ 45$^\circ$, indicating that the critical current density within the $ab$-plane is isotropic in the annealed Fe$_{1+y}$Te$_{1-x}$Se$_x$, consistent with the fourfold symmetry of the superconducting plane. Fig. 32(e) shows profiles of the magnetic induction at different temperatures along the dotted line shown in Fig. 32(d) at different temperatures. When temperature raised up to 8 K, the magnetic field can totally penetrate the sample, and a clear roof-top pattern is observed. For comparison, we also show the remanent state MO image taken on the as-grown Fe$_{1+y}$Te$_{0.6}$Se$_{0.4}$ single crystal at 10 K (see Fig. 32(f)) \cite{TaenPRB}. The white dashed lines show the sample's position. Almost no trapped field was observed except for some weak spots in the left bottom, which indicates the extremely weak (only surface or filamentary) and inhomogeneous SC of the as-grown crystal.      

\begin{figure}\center	
	\includegraphics[width=15cm]{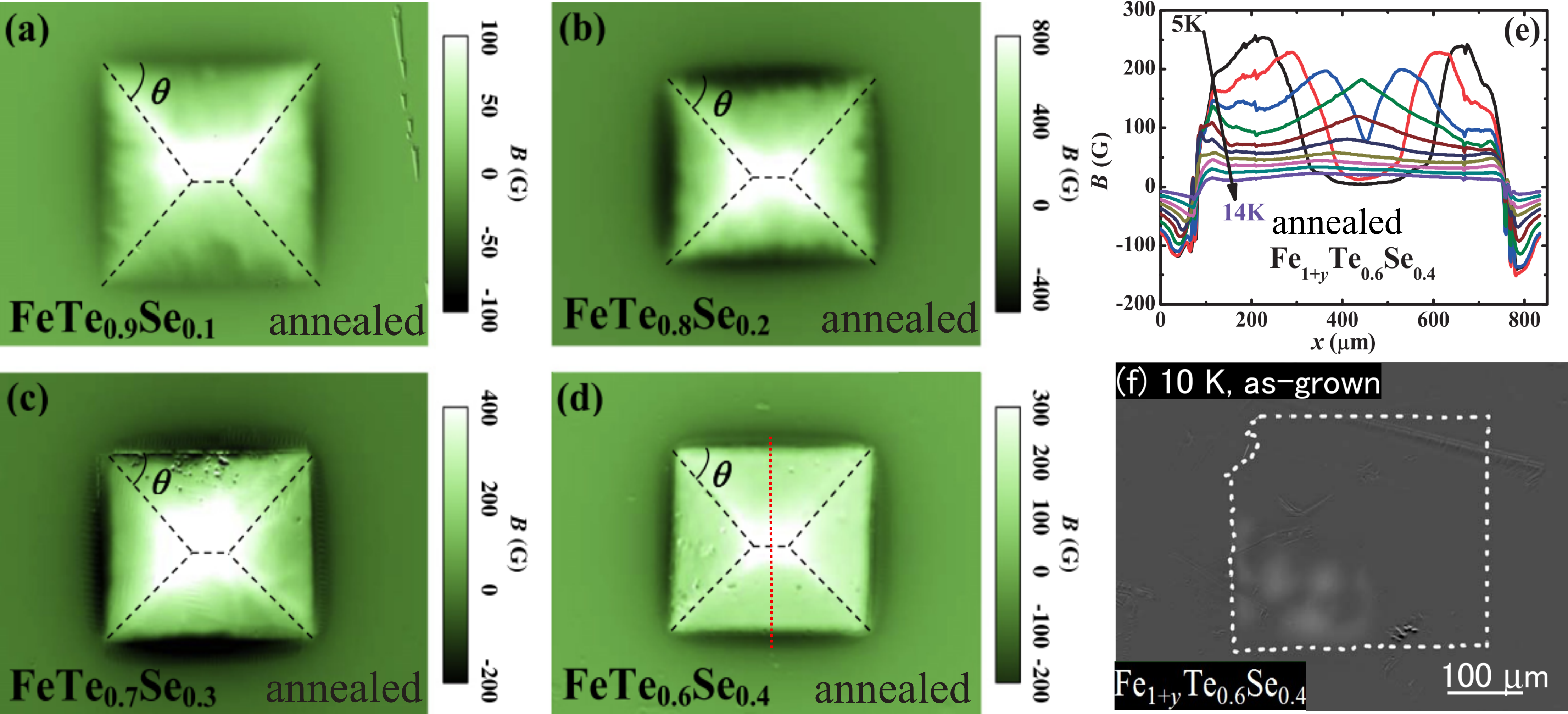}\\
	\caption{MO images in the remanent state after applying 800 Oe along the $c$-axis in annealed (a) Fe$_{1+y}$Te$_{0.9}$Se$_{0.1}$, (b) Fe$_{1+y}$Te$_{0.8}$Se$_{0.2}$, (c) Fe$_{1+y}$Te$_{0.7}$Se$_{0.3}$, and (d) Fe$_{1+y}$Te$_{0.6}$Se$_{0.4}$ \cite{SunSUSTevolution}. (e) Local magnetic induction profiles at temperatures from 5 to 14 K for Fe$_{1+y}$Te$_{0.6}$Se$_{0.4}$ taken along the dotted line in (d) \cite{SunAPRE}. (f) MO image in the remanent state of an as-grown sample at 10 K. The while dashed lines show the sample's position. (f) Reprinted with permission from \cite{TaenPRB}. Copyright 2009 by the American Physical Society.}\label{}
\end{figure} 

\subsection{Annealing effects on gap structure and superconducting pairing}

The Fe$_{1+y}$Te$_{1-x}$Se$_x$ contains a $s$-wave-like, multi-gap structure as already reported by the STM measurements \cite{HanaguriScience474}. On the other hand, the  tunneling spectra were found to be sensitive to the excess Fe as reported by Yin $et$ $al$. \cite{YinSTMNatPhy}. In the crystal free of excess Fe (Fig. 33(a)), the tunnelling spectrum exhibits a pair of sharp peaks at $\pm$1.5 meV and a pair of side peaks at $\pm$2.5 meV, together with a stateless region between the $\pm$1.5 meV peaks (Fig. 33(b)). On the other hand, in the crystal of excess Fe (Fig. 33(c) and (e)), the superconducting coherent peaks become weaker and the gap-bottom is lifted up in the spectrum taken away from the excess Fe, while the spectrum taken at the excess Fe shows a sharp peak precisely at the zero bias (Fig. 33(d)). In addition, it is also claimed that the magnitude of superconducting gap seems to be unaffected by the excess Fe, and hence the excess Fe weakens only the superconducting phase coherence, but not the strength of the superconducting pairs. Although the nodeless gap structure of Fe$_{1+y}$Te$_{1-x}$Se$_x$ has been confirmed by many experiments, the isotropic \cite{MiaoPRB,ZhangARPESScience} or anisotropic gap \cite{OkazakiARPESPhysRevLett.109.237011,ZengNatCommun} structure are still under debate. Is the controversy coming from the excess Fe need to be clarified in the future by the angle-resolved technique on crystals with (as-grown) and without (annealed) excess Fe.

Recently, the gap structure of FeSe was also reported to be affected by the Fe nonstoichiometry \cite{HopePhysRevLett.117.097003,SunPhysRevB.98.064505,SunHiraFeSePhysRevB.96.140505}. However, the nonstoichiometry in FeSe was proved to be the Fe vacancy \cite{LinWenzhiPhysRevB.91.060513}, which is different from the excess Fe in Fe$_{1+y}$Te$_{1-x}$Se$_x$ (usually $x \leq$ 0.5).                       

\begin{figure}\center
	\includegraphics[width=8.5cm]{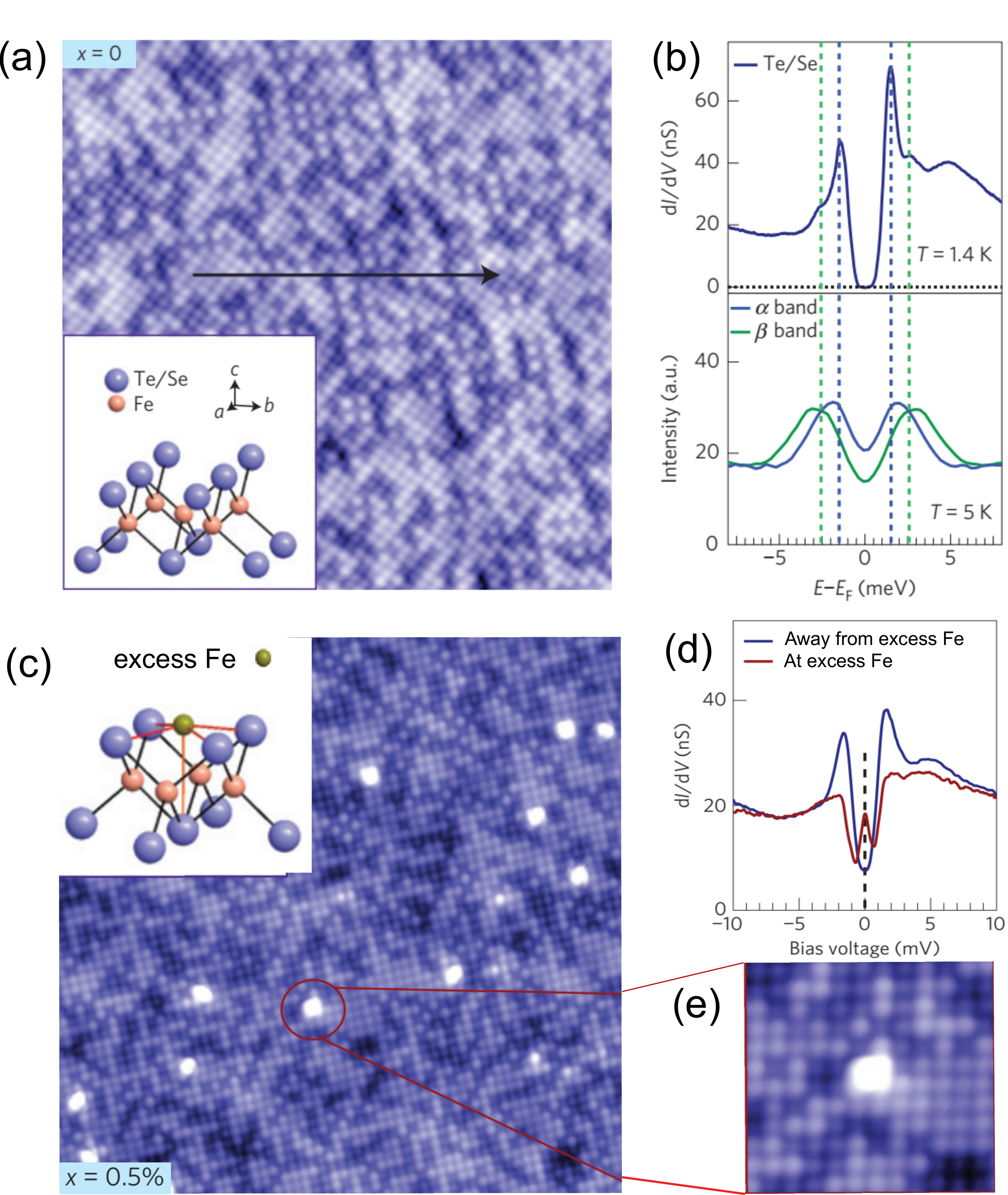}\\
	\caption{(a) Topographic image of (Te,Se) ($T$ = 1.5 K, $V$ = 1100 mV, $I$ = 0.1 nA). Inset image is the crystal structure of Fe(Te,Se) without excess Fe. (b) Comparison between STS and ARPES data. The upper panel shows STS taken on the (Te,Se) surface ($T$ = 1.4 K, $V$ = -10mV, $I$ = 0.3 nA). The lower panel shows symmetrized ARPES spectra. (c) Topographic image of (Te,Se) of the crystal with 0.5\% excess Fe (bright spots). Inset image is the crystal structure of Fe(Te,Se) with excess Fe. (d) Spectra taken at and away from the excess Fe. All the data are acquired at 1.5 K. (e) Zoomed-in image	showing one excess Fe. Reprinted with permission from \cite{YinSTMNatPhy}. Copyright 2015 by the MacMillan Publishers Ltd: Nature Physics.}\label{}
\end{figure} 

\section{Summary and perspective} 

In this topical review, we presented an overview of the reported annealing methods on Fe$_{1+y}$Te$_{1-x}$Se$_x$, and concluded that annealing in O$_2$, S, Se, Te, P, As, Sb, and I atmospheres at appropriate temperatures (200$^\circ$C - 400$^\circ$C) can totally remove the excess Fe, and introduce bulk SC. Immersing the crystals into acids or alcoholic beverages, and the electrochemical reaction method can partially remove the excess Fe, mainly on the surface layers, which will introduce weak and inhomogeneous SC. On the other hand, annealing in vacuum and N$_2$ has no effect to the excess Fe. The annealing effect of F$_2$ (provided by CaF$_2$ and SmF$_3$) and H$_2$ are also reported but need to be further tested. The annealing mechanism is that the excess Fe is attracted to the surface and reacts with the atmosphere-elements, and forms the FeM$_x$ surface layers. The annealing effect mainly evolves from the edge of the crystal to the center. We systematically compared the normal state and superconducting properties of the crystals before (with excess Fe) and after annealing (without excess Fe). Almost all the studied properties including the crystal structure, magnetism, transport properties, band structure, $T_{\rm{c}}$, phase diagram, upper critical field, anisotropy, critical current density, gap structure, and superconducting pairing are all affected by the excess Fe. Thus, we strongly suggest that future study on Fe$_{1+y}$Te$_{1-x}$Se$_x$ should pay attention to the Fe content of the used crystals.

The nonstoichiometric Fe seems to be a common issue for the IBSs-11 system. In FeSe and the S-doped one, the nonstoichiometry was found to be Fe vacancy rather than excess Fe. The annealing may be still helpful to prepare stoichiometric FeSe, which is preferred for probing its the SC and nematicity. In the S-doped FeTe compounds, the excess Fe was also confirmed. Although there are some reports about the annealing effect and the improvement of SC similar to the Fe$_{1+y}$Te$_{1-x}$Se$_x$ \cite{SunSUST,DeguchiAlcoholicSUST,MizuguchiEPL,YamamotoFeTeSJPSJ}, there are still no evidence for bulk superconductivity and total removal of excess Fe, and the SC of Fe$_{1+y}$Te$_{1-x}$S$_x$ is still mainly on the surface. The effect of excess Fe seems to be stronger in the interstitial site of the Te/S layer than the Te/Se layer, which makes the annealing more difficult. The Fe$_{1+y}$Te$_{1-x}$S$_x$ may be another interesting system similar to the sister compounds Fe$_{1+y}$Te$_{1-x}$Se$_x$ and FeSe. However, it has not been well studied due to the lack of high-quality single crystals. Appropriate annealing in the future to totally remove the excess Fe may open the door to the treasure of Fe$_{1+y}$Te$_{1-x}$S$_x$.              
           
\section{Acknowledgments}

The authors would like to thank Y. Tsuchiya, T. Taen, T. Yamada, S. Pyon, W. Zhou, X. Li, J. T. Chen, A. Sugimoto, T. Ekino, T. Nishizaki, Y. Q. Pan, and S. Ooi for their help in experiments. They would also like to thank P. Zhang, H. Kitano, and T. Machida for helpful discussions. The present work was partly supported by the Strategic Priority Research Program of Chinese Acsdemy of Science (XDB25000000), National Natural Science Foundation of China (11674054, 11611140101), and KAKENHI (19K14661, 17H01141) from JSPS.

\section*{References}
\bibliographystyle{iopart-num}
\bibliography{references}
\end{document}